\documentclass[%
 reprint,
superscriptaddress,
showpacs,
 amsmath,amssymb,
aps,
prb,
floatfix,
]{revtex4-1}

\usepackage{graphicx}
\usepackage{dcolumn}
\usepackage{bm}
\usepackage{paralist}
\usepackage{gensymb}
\usepackage{color}
\usepackage{multirow}

\begin{document}

\newcommand{\eg}{\textit{e.g.}~}
\newcommand{\cf}{\textit{cf.}~}
\newcommand{\etal}{\textit{et al.}~}
\newcommand{\ie}{\textit{i.e.} }
\newcommand{\refpar}[1]{(\ref{#1})}
\newcommand{\hatbfr}{\hat{\mathbf{r}}}
\newcommand{\hatR}{\hat{R}}
\newcommand{\diffd}{\mathrm{d}}
\newcommand{\perse}{\textit{per se}}

\newcommand\T{\rule{0pt}{2.6ex}}       
\newcommand\B{\rule[-1.2ex]{0pt}{0pt}} 

\def\csg#1{{\tt\color{red}{#1}}}


\title{On representing chemical environments}

\author{Albert P. Bart\'ok}
 \email{ab686@cam.ac.uk}
\affiliation{
Department of Engineering, University of Cambridge, Trumpington Street, Cambridge, CB2 1PZ, United Kingdom}
\author{Risi Kondor}
\affiliation{
Department of Computer Science, University of Chicago, 1100 East 58th Street, Chicago, IL 60637, United States of America}
\author{G\'abor Cs\'anyi}
\affiliation{
Department of Engineering, University of Cambridge, Trumpington Street, Cambridge, CB2 1PZ, United Kingdom}

\date{\today}

\begin{abstract}
We review some recently published methods to represent atomic neighbourhood environments, and analyse their relative merits in terms of their faithfulness and suitability for fitting potential energy surfaces. The crucial properties that such representations (sometimes called \emph{descriptors}) must have are differentiability with respect to moving the atoms, and invariance to the basic symmetries of physics: rotation, reflection, translation, and permutation of  atoms of the same species.  We demonstrate that certain widely used descriptors that initially look quite different are specific cases of a  general approach, in which a finite set of basis functions with increasing angular wave numbers are used to expand the atomic neighbourhood density function. Using the example system of small clusters, we quantitatively show that this expansion needs to be carried to higher and higher wave numbers as the number of neighbours increases in order to obtain a faithful representation, and that variants of the descriptors converge at very different rates.
We also propose an altogether new approach, called Smooth Overlap of Atomic
Positions (SOAP), that sidesteps these difficulties by directly defining the similarity between any two neighbourhood environments, and show that it is still closely connected to the invariant descriptors.
We test the performance of the various representations by fitting models to the potential energy surface of small silicon clusters and the bulk crystal. 

\end{abstract}

\maketitle

\section{Introduction}

The appropriate representation of atomic environments is a crucial ingredient of algorithms used in modern computational chemistry and condensed matter physics. For example, in structure search applications\cite{Pickard:2008ty}, each configuration depends numerically on the precise initial conditions and the path of the search, so it is important to be able to identify equivalent structures and detect similarities. In molecular dynamics simulations of
phase transitions\cite{Wales:2004vn}, one needs good order parameters that are capable of detecting changes in the local order around atoms. Typically the representation is in terms of a descriptor (also called a fingerprint), a tuple of real valued functions of the atomic positions,  \eg bond lengths, bond angles, etc.  ``In silico'' drug discovery\cite{Obrezanova:2007fz,Segall:2011ke} and other areas of chemical informatics also rely on characterising molecules using descriptors.  When constructing interatomic potentials and fitting potential energy surfaces (PES)\cite{Behler:2007th,Raff:2012vr,Braams:2009tl,Bartok:2010fj}, the driving application behind the present work, the functional forms depend on components of a carefully chosen representation of atomic neighbourhoods.

While specifying the position of each atom in a Cartesian coordinate system provides a simple and unequivocal description of atomic configurations, it is not directly suitable for making comparisons between structures: the list of coordinates is ordered arbitrarily and two structures might be mapped to each other by a rotation, reflection or translation so that two different lists of atomic coordinates can, in fact, represent the same or very similar structures. A good representation is \emph{invariant} with respect to permutational, rotational, reflectional and translational symmetries, while retaining the faithfulness of the Cartesian representation. 
In particular, a system of invariant descriptors $q_1,q_2,\ldots,q_M$ is said to be \emph{complete} if it uniquely determines the atomic environment, up to symmetries. It is said to be \emph{over-complete} if it contains spurious descriptors in the sense that a proper subset of $\{q_1,q_2,\ldots,q_M\}$ is, by itself, complete. If a representation is \emph{complete}, then there is a one-to-one mapping (i.e. a bijection) between the genuinely different atomic environments and the
invariant tuples comprising the representation. An \emph{over-complete} representation assigns potentially many distinct descriptors to a given atomic structure, but guarantees that genuinely different atomic structures will never have identical descriptors associated with them: the function relating representations to atomic structures is a surjection.

Fitting potential energy surfaces (PESs) and electrostatic multipole moment surfaces of small molecules to data generated by first principles electronic structure calculations has been a mainstay of computational chemistry for decades \cite{Tromans:1998ve,Ischtwan:1994dj,Collins:2002cq,Ho:1996kf,Maisuradze:2003kh,Guo:2004kn,Huang:2004ib,Zhang:2006gd,Blank:1995dy,Gassner:1998tn,Lorenz:2004eb,Brown:2003tu,Huang:2005ij,Manzhos:2006fd,Manzhos:2006jn,Handley:2009wg,Partridge:1997kh,Tipping:1971ui,Braams:2009tl,Cencek:2008dc}.  Typically, when modelling the PES of a small group of atoms, the list of pairwise distances is used or, equivalently, some transformed version of the interatomic distances, e.g. reciprocal\cite{Raff:2012vr} or exponential\cite{Braams:2009tl}. This description works when the number of atoms is fixed. Even in this case, a seemingly new configuration is obtained by just permuting the order of atoms, \ie crucial symmetries may be missing in this framework. Braams and Bowman\cite{Braams:2009tl} remedied this last shortcoming by using polynomials of pairwise distances, constructed such that each term is invariant to the permutation of identical atoms. Computer code is available that generates the permutationally invariant polynomials automatically\cite{Braams:2009tl} (up to ten atoms),  but this approach still does not allow for varying number of atoms in the database of configurations.

In order to generate interatomic potentials for solids or large clusters, capable of describing a wide variety of conditions, the number of neighbours that contribute to the energetics of an atom has to be allowed to vary, with the symmetry invariant descriptors remaining continuous and differentiable. Even though it is possible to allow the dimensionality $M$ to change with the number of neighbours, for the purpose of function fitting it is more practical that $M$ remains the same. None of the traditional representations fulfil this criterion. Recently, however, a number of new, promising descriptors have been proposed together with potential energy surfaces  based on them \cite{Behler:2007th,ESanville:2008jt,Bartok:2010fj,Handley:2009wg,Rupp:2012kx}. Behler and Parrinello's  ``symmetry functions''\cite{Behler:2007th} were used to generate potentials for silicon\cite{Behler:2008vo}, sodium\cite{Eshet:2012dz}, zinc oxide\cite{Artrith:2011em} and water\cite{Morawietz:2012kt} amongst others; Bartok \etal employed the bispectrum \cite{Bartok:2010fj} to fit a many-body potential for crystalline phases and defects in diamond.
Sanville \etal used a subset of internal coordinates to fit silicon potentials \cite{ESanville:2008jt}. Rupp \etal used the ordered eigenvalues of the Coulomb-matrix to fit atomisation energies \cite{Rupp:2012kx} of a set of over 7000 small organic molecules. At this point it is not clear which method of representing atomic neighbourhoods will prove to be optimal in the long term. We attempt to disentangle this issue from the rather complex details of generating first principles data and fitting PES, and separately consider the problem of constructing good descriptors.

The most well-known invariants describing atomic neighbourhoods are the bond-order parameters originally proposed by Steinhardt \etal\cite{Steinhardt:1983uh} Here we show that the bond-order parameters form a subset of a more general set of invariants called the \emph{bispectrum}\cite{Kondor:2007wz}. The formally
infinite array of bispectrum components provides an over-complete system of invariants, and by truncating it one obtains representations whose sensitivity can be refined at will. We relate the bispectrum to the representation proposed by Behler \etal\cite{Behler:2007th,Behler:2011it}, and show that, together with another descriptor set described below, their angular parts are all simple polynomials of the same set of canonical invariants.

The paper is organised as follows. In section \ref{sec:pes} we briefly recall how potential energy surfaces are constructed using invariant descriptors. In section \ref{sec:descriptors} we describe a number of descriptors, starting with a simple {\em distance metric} between atomic configurations which will be used as a reference to assess the faithfulness of all other descriptors but which itself is not differentiable.
In Section \ref{sec:soap} we introduce a new, \emph{continuous and differentiable} distance metric for constructing potential energy surfaces, called Smooth Overlap of Atomic Functions (SOAP), which has superior properties.
In section \ref{sec:results} we show numerical tests that help assess the degree of completeness and faithfulness of various descriptors and SOAP, and also show an explore their performance in fitting models for small silicon clusters and the bulk crystal. 


\section{Potential energy surface fitting}\label{sec:pes}


The main motivation behind this paper is to define and assess a family of invariant descriptors to be used for fitting interatomic potentials, or potential energy surfaces (PESs). In the construction of potentials for materials applications, the short range part of the total energy is decomposed into atomic contributions,
\begin{equation*}
E_\mathrm{short} = \sum_n \varepsilon(q_1^{(n)},\ldots,q_M^{(n)})
\textrm{,}
\end{equation*}
where $\varepsilon$ is the contribution of the $n$th atom, and $\mathbf{q}^{(n)}=(q_1^{(n)},\ldots,q_M^{(n)})$ is a system of descriptors characterizing the local atomic environment.  

Traditionally, such atomic energy functions are defined in closed form.  However, recently, there has been a lot of interest in using more flexible, \emph{nonparametric} PESs, derived from computing the total energy and its derivatives at a certain 
set $\{\mathbf{q}^{(1)},\ldots,\mathbf{q}^{(N)}\}$ of ``training'' configurations using first principles calculations.  
A crucial question then is how to fit $\varepsilon$ to the computed 
datapoints. 
The simplest approach is to use a linear fit, while 
Behler and Parrniello advocate using artificial neural networks (NN)\cite{Behler:2007th}, and 
Bart\'ok {\em et al.}~use Gaussian Approximation Potentials\cite{Bartok:2010fj}. 
However, ultimately, each of these procedures result in a PES of the form 
\begin{equation}
\label{eq:GP_kernel}
\varepsilon (\mathbf{q}) = \sum_{k=1}^N \alpha_k K\bigl(\mathbf{q},\mathbf{q}^{(k)} \bigr)
\textrm{,}
\end{equation}
where the coefficient vector $\boldsymbol{\alpha}=(\alpha_1,\ldots,\alpha_N)$ is determined by the fitting procedure,
 and $K$ is a fixed (nonlinear) function, called the kernel, whose role, 
intuitively, is to capture the degree of similarity between the atomic environments described by its two arguments. 
Clearly, then, the choice of descriptors, in particular, their invariance to symmetries, as 
well as the choice of kernel, are critical ingredients to obtaining good quality PESs. 

In general, the kernel $K$ can be interpreted as a covariance function, and therefore  
it must be symmetric and positive definite (meaning that $K(\mathbf{q},\mathbf{q'})=K(\mathbf{q'},\mathbf{q})$
and for any non-zero vector $\boldsymbol{\alpha}$ of coefficients, 
$\sum_k \sum_\ell \alpha_k \alpha_\ell K(\mathbf{q}^{(k)},\mathbf{q}^{(\ell)})>0$).  
Rasmussen and Williams\cite{Williams:2007vz} present a number of such kernels, some of the simplest ones being the following. 

The dot-product (DP) kernel is defined as
\begin{equation}\label{eq:dp-kernel}
K_{\mathrm{DP}} \bigl(\mathbf{q},\mathbf{q}^\prime \bigr) = \sum_j q_j q_j^\prime
\textrm{,}
\end{equation}
which, when substituted into equation~\refpar{eq:GP_kernel}, results in
\begin{eqnarray*}
\varepsilon (\mathbf{q}) &=& \sum_{k=1}^N \alpha_k \sum_j q_j q_j^{(k)} = \sum_j q_j \sum_{k=1}^N \alpha_k q_j^{(k)} \\
&=&\sum_j q_j \beta_j \equiv \mathbf{q} \cdot \boldsymbol{\beta}
\textrm{,}
\end{eqnarray*}
\ie the linear regression on the descriptor elements with coefficient vector $\boldsymbol{\beta}$.

When using artificial neural networks with $N_H$ hidden units, the atomic energy function $\varepsilon$ is given by 
\begin{equation}
\label{eq:NN}
\varepsilon (\mathbf{q}) = b + \sum_{j=1} ^{N_H} v_j h(\mathbf{q},\mathbf{u}_j)
\textrm{,}
\end{equation}
where $b$ is the bias, $\mathbf{v}$ the vector of unit weights, $h$ is the transfer function, and $\{\mathbf{u}_j\}_{j=1}^{N_H}$ the unit parameters\cite{Williams:2007vz}. In the limit of an infinite number of hidden units, for specific transfer functions it is possible to reformulate equation~\refpar{eq:NN} in the form of equation~\refpar{eq:GP_kernel} with well-defined covariance functions\cite{Mackay:tw,Williams:2007vz,Neal:1995ws}. For example, for $h(\mathbf{q},\mathbf{u}) = \tanh(u_0 + \sum u_j q_j)$ the corresponding kernel is\cite{Neal:1995ws}
\begin{gather*}
K_{\mathrm{NN}}\bigl(\mathbf{q},\mathbf{q}^\prime \bigr) \sim  - | \mathbf{q} - \mathbf{q}^\prime \bigr |^2 + \textrm{const.}
\end{gather*}

Finally, the squared exponential (SE) kernel that we have used in the past with the Gaussian Approximation Potentials\cite{Bartok:2010fj} and in some of the examples below is
\begin{equation}\label{eq:sekernel}
K_{\mathrm{SE}}\bigl(\mathbf{q},\mathbf{q}^\prime \bigr) = \exp \Bigl( -\sum_j \frac{(q-q^\prime)^2}{2\sigma_j^2} \Bigr)
\textrm{.}
\end{equation}


\section{Descriptors}\label{sec:descriptors}

Among the applications mentioned in the introduction, some require representing the geometry of an entire molecule, while for others one needs to describe the neighbourhood of an atom perhaps within a finite cutoff distance. While these two cases are closely related, the descriptors for one are not directly suitable for the other, although often the same idea can be used to derive representations for either case. Below, we focus on representing the neighbour environment of a single atom, but for some cases we briefly mention easy generalisations that yield global molecular descriptors.

For $N$ neighbouring atomic position vectors $\{\mathbf{r}_{1}, \mathbf{r}_{2},\ldots,\mathbf{r}_{N}\}$ taken relative to a central atom,  the symmetric matrix
\begin{equation}\label{eq:rirj}
\Sigma=\left[
\begin{array}{cccc}
\mathbf{r}_{1} \cdot \mathbf{r}_{1} & \mathbf{r}_{1} \cdot \mathbf{r}_{2} & \cdots & \mathbf{r}_{1} \cdot \mathbf{r}_{N} \\
\mathbf{r}_{2} \cdot \mathbf{r}_{1} & \mathbf{r}_{2} \cdot \mathbf{r}_{2} & \cdots & \mathbf{r}_{2} \cdot \mathbf{r}_{N} \\
\vdots                                  & \vdots                                    & \ddots & \vdots                                    \\
\mathbf{r}_{N} \cdot \mathbf{r}_{1} & \mathbf{r}_{N} \cdot \mathbf{r}_{2} & \cdots & \mathbf{r}_{N} \cdot \mathbf{r}_{N} \\
\end{array}
\right]
\end{equation}
is, according to Weyl\cite{Weyl:1946wd},  an over-complete array of basic invariants with respect to rotation, reflection and translation. However, $\Sigma$ is not a suitable descriptor, because permutations of atoms change the order of rows and columns. For example, swapping atoms 1 and 2 results in the transformed matrix

\begin{equation}
\left[
\begin{array}{cccc}
\mathbf{r}_{2} \cdot \mathbf{r}_{2} & \mathbf{r}_{2} \cdot \mathbf{r}_{1} & \cdots & \mathbf{r}_{2} \cdot \mathbf{r}_{N} \\
\mathbf{r}_{1} \cdot \mathbf{r}_{2} & \mathbf{r}_{1} \cdot \mathbf{r}_{1} & \cdots & \mathbf{r}_{1} \cdot \mathbf{r}_{N} \\
\vdots                                   & \vdots                                    & \ddots & \vdots                                    \\
\mathbf{r}_{N} \cdot \mathbf{r}_{2} & \mathbf{r}_{N} \cdot \mathbf{r}_{1} & \cdots & \mathbf{r}_{N} \cdot \mathbf{r}_{N} \\
\end{array}
\right]
\textrm{.}
\end{equation}

To compare two structures using their Weyl matrices $\Sigma$ and $\Sigma'$,  we define a \emph{reference} distance metric
\begin{equation}\label{eq:d_refp}
d_\textrm{ref} = \min_\mathbf{P} || \Sigma - \mathbf{P}\Sigma'\mathbf{P}^T ||,
\end{equation}
where $\mathbf{P}$ is a permutation matrix and the minimisation is over all possible permutations. This metric is not differentiable at locations where the permutation that minimises \refpar{eq:d_refp} changes. It would also be intractable to calculate exactly for  large numbers of atoms, but nevertheless we will use this metric to assess the faithfulness of other representations for a small system. Other, differentiable invariants shown later in this paper are, however, closely related to the elements of $\Sigma$.

One way to generate permutationally invariant differentiable functions of the Weyl matrix is to compute its eigenvalues; indeed, a very similar descriptor was recently used to fit the atomisation energies of a large set of molecules\cite{Rupp:2012kx}. However, the list of eigenvalues is very far from being complete, since there are only $N$ eigenvalues, whereas the dimensionality of the configuration space of $N$ neighbour atoms is $3N-3$  (after the rotational symmetries are removed). It is also unclear how to make the descriptors based on the eigenspectrum continuous and differentiable as the number of neighbours varies. 

The Z-matrix, or internal coordinates, is a customary set of rotationally invariant descriptors usually used to describe the geometry of entire molecules, but it is not invariant to permutations of atoms. The Z-matrix is  complete, but in contrast to the Weyl matrix of basic invariants which are based solely on bond lengths and bond angles, it is a minimal set of descriptors that also contains some dihedral angles.

Another straightforward way to compare structures is based on pairing the atoms from each and finding the optimal rotation that brings the two structures into as close an alignment as possible. For each pair of structures $\{\mathbf{r}_{i}\}_{i=1}^N$ and $\{\mathbf{r}'_{i}\}_{i=1}^N$, it is possible to order the atoms according to their distance from the central atom (or centre of mass, in case we want to compare entire molecules) and compute
\begin{equation*}
\Delta(\hat R)= \sum_j^N | \mathbf{r}_{i} - \hat R \mathbf{r}'_{i} |^2
\textrm{,}
\end{equation*}
where $\hat R$ represents an arbitrary rotation (including the possibility of a reflection). We can then define the distance between two configurations as 
\begin{equation}\label{eq:d_refr}
\Delta = \min_{\hat R} \Delta(\hat R)
\textrm.
\end{equation}
This distance clearly has all the necessary invariances and completeness properties, but, like $d_\textrm{ref}$, it is not suitable for parametrising potential energy surfaces, because it is again not differentiable: the reordering procedure and the minimisation over rotations and reflections introduce cusps. 

In the field of molecular informatics, one popular descriptor is based on the histogram of pairwise atomic distances\cite{Swamidass:2005jx}, similar to Valle's crystal
fingerprint\cite{Valle:2010jt}. We will not consider these here, because they are unsuitable for fitting PES, as they are clearly not complete: \eg from six unordered distance values it is not necessarily possible to construct a unique tetrahedron, even though the number of degrees of freedom is also six.
\begin{figure}[h]
  \includegraphics[width=4.cm]{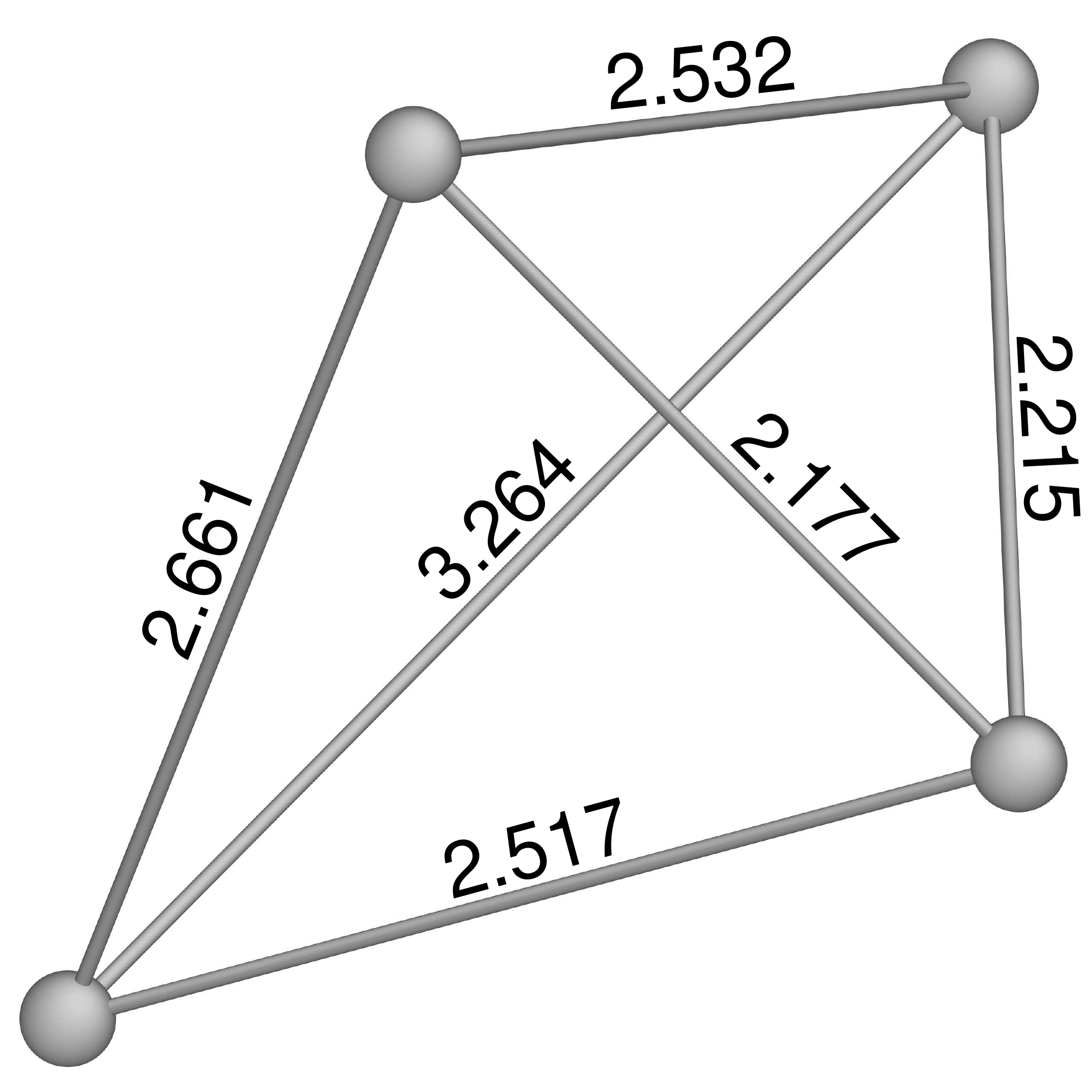}
  \includegraphics[width=4.cm]{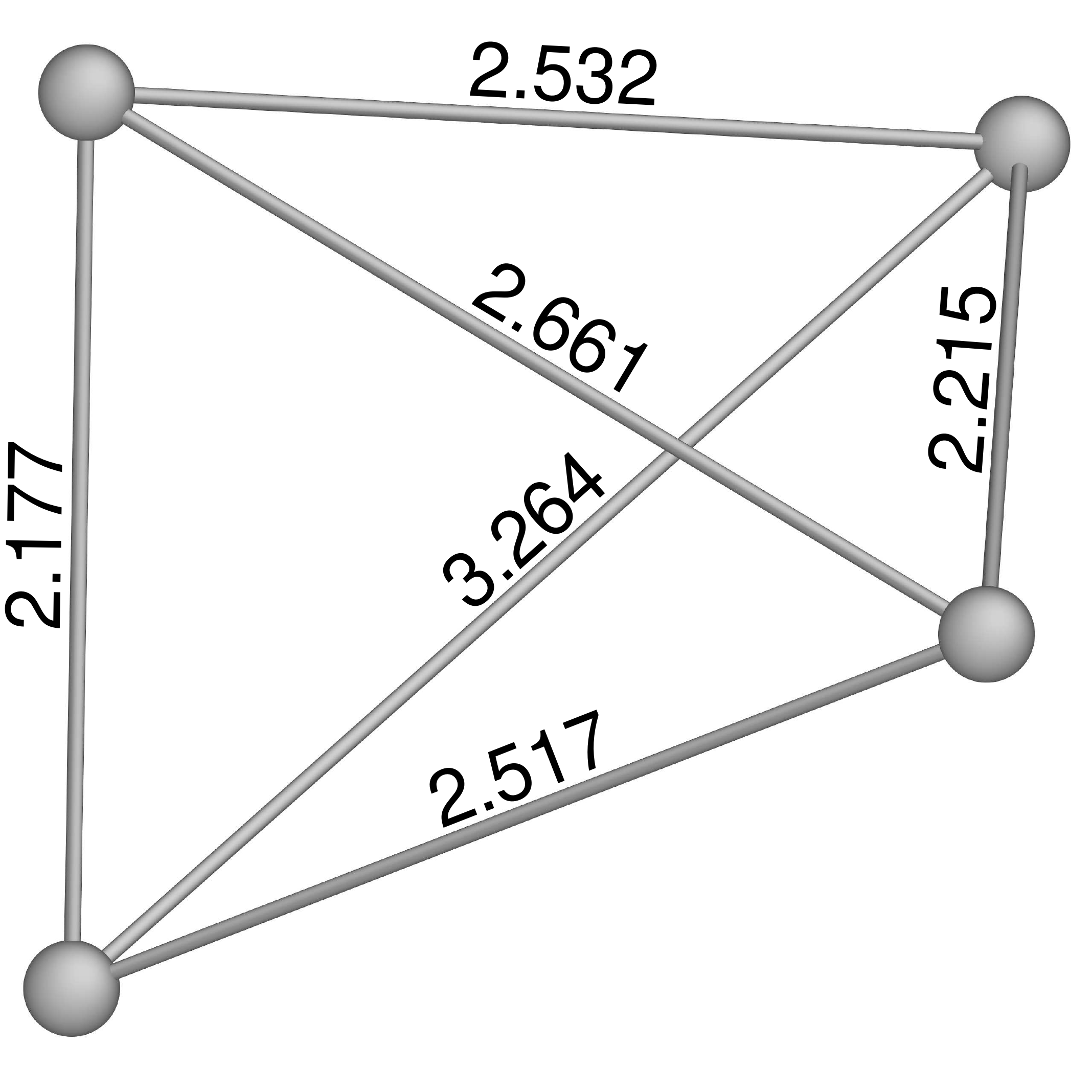}
  \caption{Two distinct tetrahedra, constructed from the same set of 6 distances.}
  \label{fgr:tetra12}
\end{figure}
Figure~\ref{fgr:tetra12}  shows two tetrahedra that were constructed such that the edges in each correspond to the same set of six distances. The tetrahedra are manifestly different, which can also be seen by comparing the lists of angles,  shown in Table~\ref{tbl:tetra12}.
\begin{table*}[hbt]
\small
\caption{Angles of the tetrahedra shown in figure~\ref{fgr:tetra12}.}
\label{tbl:tetra12}
\begin{ruledtabular}
  \begin{tabular}{dddddddddddd}
42.66\degree & 49.32\degree & 49.63\degree & 50.36\degree & 52.84\degree & 54.10\degree & 55.50\degree & 61.74\degree & 68.63\degree & 70.40\degree & 77.84\degree & 86.98\degree \\
41.78\degree & 42.66\degree & 49.63\degree & 50.36\degree & 50.42\degree & 50.80\degree & 61.74\degree & 61.77\degree & 67.81\degree & 68.63\degree & 86.98\degree & 87.42\degree \\
  \end{tabular}
\end{ruledtabular}
\end{table*}

\subsection{Bond-order parameters}

As a first step in deriving continuous invariant representations of atomic environments, we define the  atomic neighbour density function associated with a given atom as
\begin{equation}\label{eq:atomic_density}
\rho (\mathbf{r}) = \sum_i w_{Z_i}\delta(\mathbf{r}-\mathbf{r}_{i}) \textrm{,}
\end{equation}
where the index $i$ runs over the neighbours of the atom within some cutoff distance, $w_{Z_i}$ is a unique weight factor assigned according to the atomic species of $i$, and $\mathbf{r}_{i}$ is the vector from the central atom to neighbour $i$. For clarity, we will omit the species weights from now on, and assume a single atomic species, but none of our results rely on this. Determining which neighbours to include in the summation can be done by using a simple binary valued, or a smooth real valued cutoff function of the interatomic distance, or via a more sophisticated procedure, \eg Voronoi analysis\cite{Steinhardt:1983uh}. The  atomic neighbour density is already
invariant to permuting neighbours, because changing the order of the atoms in the neighbour list only affects the order of the summation. To simplify the following derivation, for now we omit the information on the radial distance to the neighbours, but will show later how the radial information can be included. The  atomic neighbour density function can then be expanded in terms of spherical harmonics:
\begin{equation}
\label{eq:rho_expand}
\rho(\hat{\mathbf{r}}) = \sum_{l=0}^\infty \sum_{m=-l}^l c_{lm}
Y_{lm} (\hat{\mathbf{r}})  \textrm{,}
\end{equation}
where $\hat{\mathbf{r}}$ is the point on the unit sphere corresponding to the direction of the vector $\mathbf{r}$, thus $\rho(\hat{\mathbf{r}})$ is the projection of $\rho(\mathbf{r})$ onto the unit sphere, $S^2$. 

The properties of functions defined on the unit sphere are related to the group theory of SO(3), the group of three dimensional rotations. Spherical harmonics form an orthonormal basis set for $L_2(S^2)$, the class of square integrable functions on the sphere:
\begin{equation*}
\langle Y_{lm} | Y_{l'm'} \rangle = \delta_{ll'} \delta_{mm'} \textrm{,}
\end{equation*}
where the inner product of functions $f$ and $g$ is defined as
\begin{equation*}
\langle f | g \rangle = \int f^*(\hat{\mathbf{r}}) g(\hat{\mathbf{r}}) \: \mathrm{d}\Omega(\hat{\mathbf{r}})
\textrm{,}
\end{equation*}
where the surface element $\mathrm{d}\Omega(\hat{\mathbf{r}})$ can be expressed in terms of the polar angles $\theta$ and $\phi$ as
\begin{equation*}
\mathrm{d}\Omega(\hat{\mathbf{r}})=\sin \theta \: \mathrm{d}\theta \: \mathrm{d}\phi
\textrm{,}
\end{equation*}
and the coefficients $c_{lm}$ are given by 
\begin{equation}
c_{lm} = \langle \rho | Y_{lm} \rangle = \sum_i
Y_{lm} (\hat{\mathbf{r}}_{i})\textrm{.} 
\label{clmdef}
\end{equation}

The quantities $Q_{lm}$ introduced by Steinhardt \etal\cite{Steinhardt:1983uh} are
proportional to the coefficients $c_{lm}$.  Dividing by $N$, the number of neighbours of the atom (within a finite cut-off distance) provides the atomic order parameters
\begin{equation}
Q_{lm} = \frac{1}{N} \sum_i Y_{lm} (\hat{\mathbf{r}}_{i}) \textrm{.}
\label{Qlmdef}
\end{equation}
Furthermore, averaging \refpar{Qlmdef} over atoms in the entire
system gives a set of global order parameters
\begin{equation*}
\bar{Q}_{lm} = \frac{1}{N_b}\sum_{ii^\prime} Y_{lm} (\hat{\mathbf{r}}_{ii^\prime}) \textrm{,}
\end{equation*}
where $N_b$ is the total number of atom pairs included in the summation, and we wrote $\mathbf{r}_{ii^\prime}$ for the vector connecting atom $i$ to its neighbour $i^\prime$. Both sets are invariant to permutations of atoms and
translations, but still depend on the orientation of the reference frame. However, rotationally
invariant combinations can be constructed as
\begin{align}
Q_l &= \left[ \frac{4 \pi}{2l+1} \sum_{m=-l}^l (Q_{lm})^* Q_{lm} \right]^{1/2}
\textrm{and} \label{Qldef} \\
W_l &= \sum_{m_1, m_2, m_3 = -l}^l \left(
\begin{array}{ccc}
l   & l   & l \\
m_1 & m_2 & m_3
\end{array} \right) Q_{lm_1} Q_{lm_2} Q_{lm_2}\label{Wldef}
\end{align}
for atomic neighbourhoods and
\begin{align*}
\bar{Q}_l &= \left[ \frac{4 \pi}{2l+1} \sum_{m=-l}^l \bar{Q}_{lm}^* \bar{Q}_{lm} \right]^{1/2} \\
\bar{W}_l &= \sum_{m_1, m_2, m_3 = -l}^l \left(
\begin{array}{ccc}
l   & l   & l \\
m_1 & m_2 & m_3
\end{array} \right) \bar{Q}_{lm_1} \bar{Q}_{lm_2} \bar{Q}_{lm_2}
\end{align*}
for the entire structure. The factor in parentheses is the Wigner-3$jm$ symbol\cite{Varshalovich:1987ul}, which is zero unless $m_1+m_2+m_3=0$.

The numbers $Q_l$ and $W_l$ are called second-order and third-order bond-order parameters, respectively. It is possible to normalise $W_l$ such that it does not depend strongly
on the number of neighbours:
\begin{equation*}
\widehat{W}_l = \left[ \sum_{m=-l}^l (Q_{lm})^* Q_{lm} \right]^{-3/2} W_l 
\textrm{.}
\end{equation*}

For symmetry reasons, only coefficients with $l \ge 4$ have non-zero values in environments with cubic
symmetry and $l \ge 6$ for environments with icosahedral symmetry. Different values correspond to crystalline materials with different symmetry, while the global order parameters vanish in disordered phases, such as liquids. Bond-order parameters were originally introduced for studying order in liquids and glasses\cite{Steinhardt:1983uh}, but were soon adopted for a wide range of applications. They have been used to study the free energy of clusters \cite{Doye:1999jn,Partay:2010cy}, melting of quantum solids \cite{Chakravarty:2000ek}, nucleation\cite{Mountain:1984il}, as well as to serve as reaction coordinates in simulations of phase transitions\cite{Duijneveldt:1992tg,Hernandez:2007ta} and glasses\cite{vanBlaaderen:1995th}, and also to generate interatomic potentials \cite{Biswas:1987vp}.

\subsection{The power spectrum}

Using some basic concepts from representation theory, we now prove that the second-order bond-order parameters are indeed rotationally invariant, then we show a more general set of third order invariants,\cite{Kondor:2007wz} of which the $Q$s and the $W$s are a subset. An arbitrary rotation $\hat{R}$ operating on a spherical harmonic function $Y_{lm}$ transforms it into a linear combination of spherical harmonics with the same $l$ index:
\begin{equation*}
\hat{R} \, Y_{lm} = \sum_{m'=-l}^l D^{l}_{mm'}(\hat R) Y_{lm'} \textrm{,}
\end{equation*}
where the $\mathbf{D}^{l}(\hat R)$ matrices are known as the Wigner matrices, which form the irreducible representations of the three dimensional rotation group, SO(3). The elements of the
Wigner matrices are given by
\begin{equation}\label{eq:wigner}
D^{l}_{mm'}(\hat R) = \langle Y_{lm} | \hat{R} | Y_{lm'} \rangle \textrm{.}
\end{equation}
It follows that the rotation operator $\hat{R}$ acts on the function $\rho$ as
\begin{align*}
\hat{R} \rho &= \hat{R} \sum_{l=0} \sum_{m=-l}^l c_{lm} Y_{lm} = \sum_{l=0} \sum_{m=-l}^l c_{lm} \hat{R} Y_{lm} \\
             &= \sum_{l=0} \sum_{m=-l}^l \sum_{m'=-l}^l c_{lm} D^{l}_{mm'}(R) Y_{lm'}\\
             &\equiv \sum_{l=0} \sum_{m'=-l}^l c'_{lm} Y_{lm'} \textrm{,}
\end{align*}
thus the column vector $\mathbf{c}_l$ of expansion coefficients transforms under rotation as
\begin{equation*}
\mathbf{c}_l \to \mathbf{D}^{l}(\hat R) \mathbf{c}_l \textrm{.}
\end{equation*}

Making use of the fact that rotations are unitary operations on $L_2(S^2)$, it is possible to show that the
matrices $\mathbf{D}^{l}$ are unitary, 
\begin{equation*}
{\mathbf{D}^l}^ \dagger \mathbf{D}^{l} = \mathbf{I} \textrm{,}
\end{equation*}
and therefore $\mathbf{c}_l^\dagger \mathbf{c}_l$ transforms according to
\begin{equation}\label{eq:power_spectrum}
p_l \equiv \mathbf{c}_l^\dagger \mathbf{c}_l \to \left( \mathbf{c}_l ^\dagger {\mathbf{D}^l}^ \dagger \right) \left( \mathbf{D}^{l} \mathbf{c}_l \right) = \mathbf{c}_l^\dagger \mathbf{c}_l
\textrm{,}
\end{equation}
i.e. is invariant under rotation. We call $p_l$ the rotational power spectrum due to the analogy with the familiar power spectrum of ordinary Fourier analysis.

We also note that the elements of $\mathbf{c}_l$ transform under reflection about the origin as
\begin{equation}\label{eq:c_reflect}
\mathbf{c}_l \to (-1)^l \: \mathbf{c}_l
\textrm{,}
\end{equation}
thus the power spectrum is also invariant to this symmetry operation.
A comparison with equations \refpar{clmdef}, \refpar{Qlmdef} and \refpar{Qldef} shows that the second-order bond-order parameters are related to the power spectrum via the simple scaling
\begin{equation*}
Q_l = \left( \frac{4 \pi}{2l+1} p_l \right)^{1/2} \textrm{.}
\end{equation*}

The power spectrum is clearly not a complete descriptor for a general function $ f(\hat{\mathbf{r}}) $ on the sphere, for example consider the two different functions
\begin{equation*}
f_1 = Y_{22} + Y_{2-2} + Y_{33} + Y_{3-3}
\end{equation*}
and
\begin{equation*}
f_2 = Y_{21} + Y_{2-1} + Y_{32} + Y_{3-2}
\textrm{,}
\end{equation*}
which both have the same power spectrum, $p_2=2$ and $p_3=2$ (with all other components equal to zero). However, for the restricted class of functions which are sums of a limited number of delta functions (such as the  atomic neighbour density $\rho$ in equation~\refpar{eq:atomic_density}), the power spectrum elements turn out to be polynomials of the basic invariants of Weyl.  Using numerical experiments we demonstrate in section~\ref{sec:results} that for a fixed number of neighbours a certain set of power spectrum components is likely to be over-complete. 


\subsection{The bispectrum}\label{sec:bispectrum}
We generalise the concept of the power spectrum to obtain a larger set of invariants via  coupling  different angular momentum channels\cite{Kondor:2007wz,Kakarala:1992}. Let us consider the direct product $\mathbf{c}_{l_1} \otimes
\mathbf{c}_{l_2}$, which transforms under a rotation as
\begin{equation*}
\mathbf{c}_{l_1} \otimes \mathbf{c}_{l_2} \to \left( \mathbf{D}^{l_l} \otimes \mathbf{D}^{l_2}
\right) \left( \mathbf{c}_{l_1} \otimes \mathbf{c}_{l_2} \right) \textrm{.}
\end{equation*}
It follows from the representation theory of compact groups that the direct product of two irreducible
representations can be decomposed into a direct sum of irreducible representations\cite{Maschke:1898uy}. In case of the SO(3) group, the direct product of two Wigner matrices can be decomposed into a direct sum of Wigner matrices,
\begin{equation}
\label{eq:CG_series}
D^{l_1}_{m_1 m'_1} D^{l_2}_{m_2 m'_2} = \sum_{l,m,m'}
D^l_{mm'} \; \bigl( C^{l\, l_1 l_2}_{m m_1 m_2} \bigr)^* \; C^{l\, l_1 l_2}_{m' m'_1 m'_2} 
\textrm{,}
\end{equation}
where $C^{l\, l_1 l_2}_{m m_1 m_2}$ denote the Clebsch-Gordan coefficients or, using more compact notation,
\begin{equation}\label{eq:cross_wigner_compact}
\mathbf{D}^{l_l} \otimes \mathbf{D}^{l_2} = \left( \mathbf{C}^{l_1 l_2} \right)^\dagger
\left[ \bigoplus_{l=|l_1-l_2|}^{l_1+l_2} \mathbf{D}^{l} \right] \mathbf{C}^{l_1 l_2} \textrm{,}
\end{equation}
where $\mathbf{C}^{l_1 l_2}$ denote the matrices formed of the Clebsch-Gordan coefficients. These are themselves unitary, so the vector $\mathbf{C}^{l_1 l_2} \left( \mathbf{c}_{l_1}
\otimes \mathbf{c}_{l_2} \right) $ transforms as
\begin{equation}\label{eq:cg_times_direct}
\mathbf{C}^{l_1 l_2} \left( \mathbf{c}_{l_1} \otimes \mathbf{c}_{l_2} \right) \to
\left[ \bigoplus_{l=|l_1-l_2|}^{l_1+l_2} \mathbf{D}^{l} \right] \mathbf{C}^{l_1 l_2} 
\left( \mathbf{c}_{l_1} \otimes \mathbf{c}_{l_2} \right) \textrm{.}
\end{equation}
Writing out the block diagonal matrix in the square brackets as
\begin{equation*}
\left[ \bigoplus_{l=|l_1-l_2|}^{l_1+l_2} \mathbf{D}^{l} \right] \equiv
\left(
\begin{array}{cccc}
\fbox {\parbox{1.0cm}{\vrule height0.9cm depth0cm width0cm \vbox to 0.9cm{\vfill\scriptsize$ \mathbf{D}^{|l_1-l_2|}$\vfill}}}&&&\\
&\hskip-0.15cm\fbox {\parbox{1.3cm}{\vrule height1.2cm depth0cm width0cm \vbox to 1.2cm{\vfill\scriptsize$ \mathbf{D}^{|l_1-l_2|+1}$\vfill}}}&&\\
&&\ddots&\\
&&&\fbox {\parbox{1.5cm}{\vrule height1.5cm depth0cm width0cm \vbox to 1.5cm{\vfill\scriptsize\hbox to 1.5cm{\hfill$\mathbf{D}^{l_1+l_2}$\hfill}\vfill} }}\\
\end{array}\right)\textrm{,}
\end{equation*}
we see that each block selects a particular slice of the vector in equation \refpar{eq:cg_times_direct},  which transforms according to a given $\mathbf{D}^l$ matrix. We give a new symbol to these slices, $\mathbf{g}_{l\, l_1 l_2}$, so that the original vector is their direct sum
\begin{equation*}
\mathbf{C}^{l_1 l_2} \left( \mathbf{c}_{l_1} \otimes \mathbf{c}_{l_2} \right)
\equiv  \bigoplus_{l=|l_1-l_2|}^{l_1+l_2} \mathbf{g}_{l\,l_1 l_2} ,
\end{equation*}
and each $\mathbf{g}_{l\, l_1 l_2}$  transforms under rotation as
\begin{equation*}
\mathbf{g}_{l\, l_1 l_2} \to \mathbf{D}^{l} \mathbf{g}_{l\, l_1 l_2} \textrm{.}
\end{equation*}

Analogously to the power spectrum, we can now define the bispectrum as the collection of scalars
\begin{equation*}
b_{l\, l_1 l_2} = \mathbf{c}_l^\dagger \mathbf{g}_{l\, l_1 l_2} \textrm{,} 
\end{equation*}
which are invariant to rotations:
\begin{equation*}
b_{l\, l_1 l_2} =  \mathbf{c}_l^\dagger \mathbf{g}_{l\, l_1 l_2} \to 
\left( \mathbf{c}_l \mathbf{D}^{l} \right)^\dagger \mathbf{D}^{l} \mathbf{g}_{l\, l_1 l_2}
= \mathbf{c}_l^\dagger \mathbf{g}_{l\, l_1 l_2}
\textrm{.}
\end{equation*}
It follows from equation~\refpar{eq:c_reflect} that those elements of the bispectrum where $l_1+l_2+l$ is odd change sign under reflection about the origin.
If invariance to reflection is required, we take the absolute value of these components or omit them from the descriptor.

Rewriting the bispectrum formula as
\begin{equation}
\label{eq:bispectrum_so3}
b_{l\, l_1 l_2} = \sum_{m=-l}^l \sum_{m_1=-l_1}^{l_1} \sum_{m_2=-l_2}^{l_2} c_{lm}^*
C_{m m_1 m_2}^{l\,l_1 l_2} c_{l_1 m_1} c_{l_2 m_2} \textrm{,}
\end{equation}
the similarity to the third-order bond-order parameters becomes apparent.
Indeed, the Wigner
3$jm$-symbols are related to the Clebsch-Gordan coefficients through
\begin{equation}\label{eq:wigner3jm_CG}
\left(
\begin{array}{ccc}
l_1 & l_2 & l_3 \\
m_1 & m_2 & m_3
\end{array} \right) = \frac{(-1)^{l_1-l_2-m_3}}{\sqrt{2l_3+1}} C_{m_1 m_2 -m_3}^{l_1 l_2 l_3}\textrm{,}
\end{equation}
and by substituting the spherical harmonics identity $Y_{lm} = (-1)^m Y^*_{l-m}$ in equation~\refpar{Qlmdef} it follows that
\begin{equation}\label{eq:qlm_identity}
Q_{lm} = (-1)^m ( Q_{lm} )^*
\textrm{.}
\end{equation}
Substituting the identities~\refpar{eq:wigner3jm_CG} and \refpar{eq:qlm_identity} into the definition~\refpar{Wldef} we obtain
\begin{multline*}
W_l = \frac{1}{\sqrt{2l+1}} \times \\
\sum_{m_1,m_2,m_3=-l}^{l} (-1)^{-m} C_{m_1 m_2 -m_3}^{l\,l\,l} Q_{lm_1} Q_{lm_2} (-1)^m ( Q_{lm_3})^*
\textrm{,}
\end{multline*}
thus the third-order parameters $W_l$ are seen to be proportional to the diagonal elements of the bispectrum, $b_{l\,l\,l}$. Noting that $Y_{00} \equiv 1$, the coefficient $c_{00}$ is the number of neighbours $N$, and using $C_{m\,0\, m_2}^{l\, 0\, l_2} = \delta_{l\,l_2} \delta_{m m_2}$, the bispectrum elements $l_1=0$, $l=l_2$ are identical to the previously introduced power spectrum components:
\begin{align*}
b_{l\,0\,l}  &= N \sum_{m=-l}^l \sum_{m_2=-l}^{l} c_{lm}^* \delta_{m m_2} c_{l m_2} =\\
&=N \sum_{m=-l}^l
c_{lm}^* c_{lm} = N p_l \textrm{,}
\end{align*}
therefore,
\begin{align*}
Q_l &\propto \sqrt{p_l} \propto \sqrt{b_{l\,0\,l}}\\
W_l & \propto b_{l\, l\, l}
\textrm{.}
\end{align*}

The first few terms of the power spectrum and bispectrum for an atom with three neighbours are shown below, where $\theta_{ii^\prime}$ is the angle between the bonds to neighbours $i$ and $i^\prime$, and the sums are over all the neighbours.

\begin{align*}
&p_0 = \frac{9}{4\pi} \\
&p_1 = \frac{3}{4\pi} \left( \sum_{ii^\prime} \cos \theta_{ii^\prime} + 3 \right) \\
&p_2 = \frac{5}{4\pi} \left( \frac{3}{2}\sum_{ii^\prime} \cos^2 \theta_{ii^\prime} + 6 \right) \\
&p_3 = \frac{7}{4\pi} \left( \frac{5}{2}\sum_{ii^\prime} \cos^3 \theta_{ii^\prime} - \frac{3}{2}\sum_{ii^\prime} \cos \theta_{ii^\prime} + 3 \right) \\
&p_4 = \frac{9}{16\pi} \left(\frac{35}{2}\sum_{ii^\prime} \cos^4 \theta_{ii^\prime} - 15\sum_{ii^\prime} \cos^2 \theta_{ii^\prime} + 13 \right)\\
\end{align*}

\begin{align*}
&b_{211} = \sqrt{\frac{15}{128\pi^3}} \left(
\frac{3}{4} \left(\sum_{ii^\prime} \cos \theta_{ii^\prime} \right)^2 + \right. \\
&\left. + \frac{3}{2} \sum_{ii^\prime} \cos^2 \theta_{ii^\prime} +  
 5 \sum_{ii^\prime} \cos \theta_{ii^\prime}
\right) \\
&b_{321} = \frac{150}{8}\sqrt{\frac{7}{\pi^3}} \left(
\frac{5}{2} \sum_{ii^\prime} \cos^3 \theta_{ii^\prime} + \right. \\ 
& \left. + \frac{5}{4} \sum_{ii^\prime} \cos^2 \theta_{ii^\prime}  \sum_{ii^\prime} \cos \theta_{ii^\prime} -\frac{1}{2}\left(\sum_{ii^\prime} \cos \theta_{ii^\prime} \right)^2 + \right. \\
& \left. + 4 \sum_{ii^\prime} \cos^2 \theta_{ii^\prime} - 2\sum_{ii^\prime} \cos \theta_{ii^\prime} + 18
\right)
\end{align*}


\subsection{Radial basis}

Thus far we neglected the distance of neighbouring atoms from the central atom by using the unit-sphere projection of the atomic environment.  One way to introduce radial information is to complement the spherical harmonics basis in
equation~\refpar{eq:rho_expand} with radial basis
functions $g_n$\cite{Taylor:2009em}, 
\begin{equation}\label{eq:radbas}
\rho(\mathbf{r}) = \sum_n \sum_{l=0} \sum_{m=-l}^l c_{nlm} g_n(r)
Y_{lm} (\hat{\mathbf{r}}) \textrm{.}
\end{equation}
If the set of radial basis functions is not orthonormal, \ie $\langle g_n | g_m \rangle = S_{nm} \ne \delta_{nm}$, after obtaining the coefficients $c'_{nlm}$ with
\begin{equation*}
c'_{nlm} = \langle g_n Y_{lm} | \rho \rangle \textrm{,}
\end{equation*}
the elements $c_{nlm}$ are given by
\begin{equation*}
c_{nlm} = \sum_{n'} \left(S^{-1}\right)_{n'n} c'_{n'lm} \textrm{.}
\end{equation*}
In practice, when constructing the bispectrum, either $c'_{nlm}$ or $c_{nlm}$ can be used.

\begin{figure}[h]
\includegraphics[width=8.6cm]{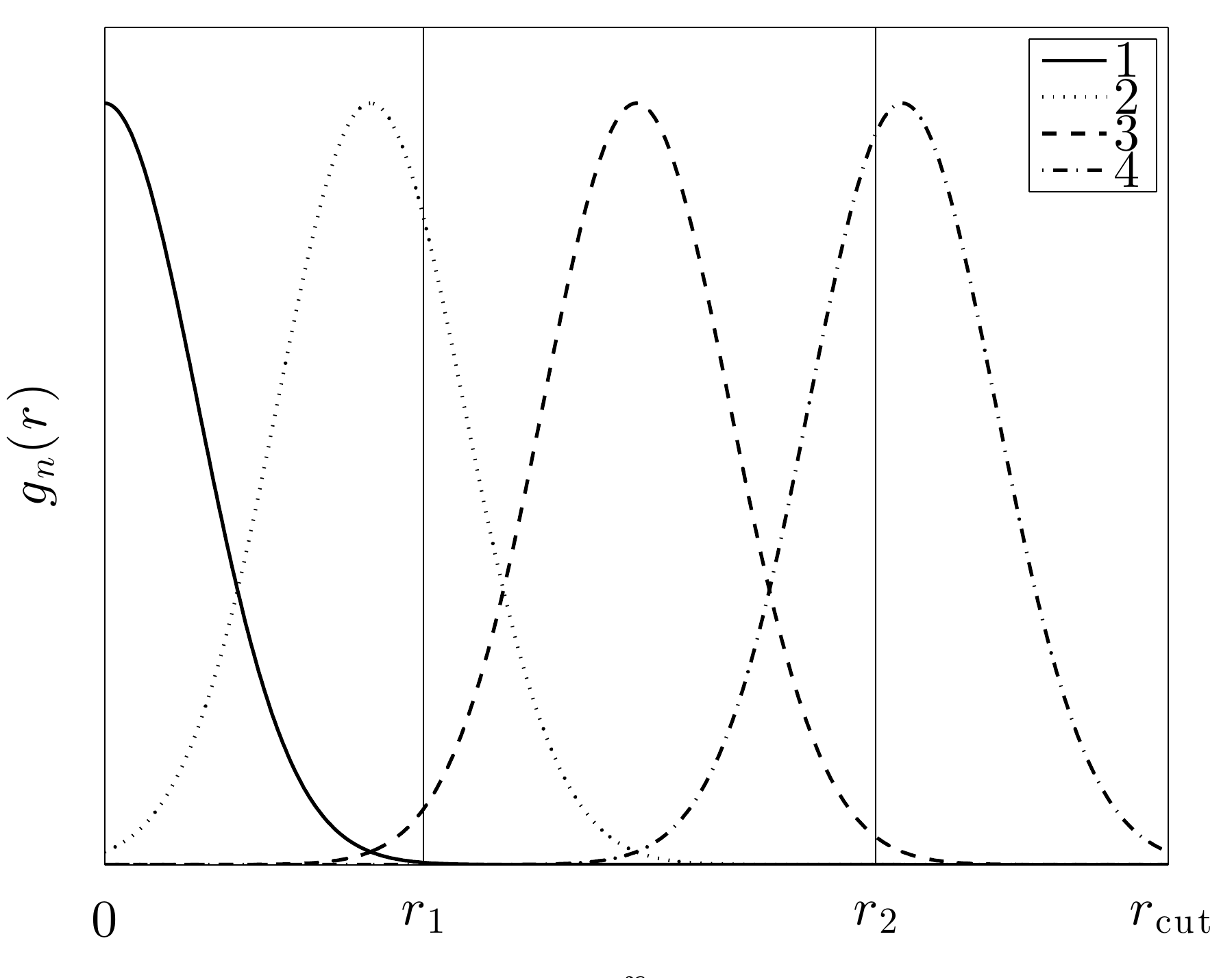}
\caption{Example of weakly overlapping radial basis functions $g_n(r)$, \cf equation~\refpar{eq:radbas}. Atoms 1 and 2 at distance $r_1$ and $r_2$ from the centre become decoupled as their contribution to the power spectrum or bispectrum elements is weighed down by the product $g_n(r_1)g_n(r_2)$, which is rather small for all $n$.}
\label{fig:radial_gauss}
\end{figure}

Rotational invariance must only apply globally, and not to each radial basis separately, therefore the angular momentum channels corresponding to different radial basis functions need to be coupled. So, although extending equations~\refpar{eq:power_spectrum} and \refpar{eq:bispectrum_so3} simply as
\begin{align*}
&p_{n l}=\sum_{m=-l}^l c_{nlm}^* c_{n l m} \\
&b_{n l\, l_1 l_2} = \sum_{m=-l}^l \sum_{m_1=-l_1}^{l_1} \sum_{m_2=-l_2}^{l_2} c_{nlm}^*
C_{m m_1 m_2}^{l\,l_1 l_2} c_{nl_1 m_1} c_{nl_2 m_2}
\end{align*}
provides a set of invariants describing the three-dimensional neighbourhood of the atom, this can easily lead to a poor representation if the radial basis functions do not sufficiently overlap. The different radial shells will only be weakly coupled, and the representation will have spurious quasi-invariances to rotating subsets of atoms at approximately the same distance, as illustrated in Figure~\ref{fig:radial_gauss}.

To avoid this, it is necessary to choose basis functions that are sensitive over a wide range of distances, although this may reduce the sensitivity of each radial basis function, because they are varying very slowly. The fine-tuning of the basis set is rather arbitrary, and there is no guarantee that a choice exists that is optimal or even satisfactory for all systems of interest.

We suggest constructing radial functions from cubic and higher order polynomials, $\phi_\alpha(r)=(r_{\mathrm{cut}} - r)^{\alpha+2}/N_\alpha$ for $\alpha=1,2, \ldots, n_\mathrm{max}$, normalised on the range $(0,r_{\mathrm{cut}})$ using
\begin{equation*}
N_\alpha =  \sqrt{ \int _0 ^{r_{\mathrm{cut}}} (r_{\mathrm{cut}} - r)^{2(\alpha+2)} \mathrm{d}r}=\sqrt{ \frac{ r_{\mathrm{cut}}^{2\alpha+5} }{2\alpha+5} }
\textrm{.}
\end{equation*}

The orthonormalised construction
\begin{equation}\label{eq:afs_radial}
g_n(r) = \sum_{\alpha=1} ^{n_\mathrm{max}}  W_{n\alpha} \phi_\alpha(r)
\end{equation}
guarantees that each radial function returns smoothly to zero at the cutoff with both the first and the second derivative equal to zero. The matrix $\mathbf{W}$ of linear combination coefficients  is obtained from the overlap matrix as

\begin{equation*}
S_{\alpha \beta} = \int _0 ^ {r_{\mathrm{cut}}} \phi_\alpha(r) \phi_\beta(r) \mathrm{d}r = \frac{\sqrt{(5+2\alpha)(5+2\beta)}}{5+\alpha+\beta}
\end{equation*}

\begin{equation*}
\mathbf{W} = \mathbf{S}^{-1/2}
\textrm{.}
\end{equation*}

\begin{figure}[htb]
\includegraphics[width=8.6cm]{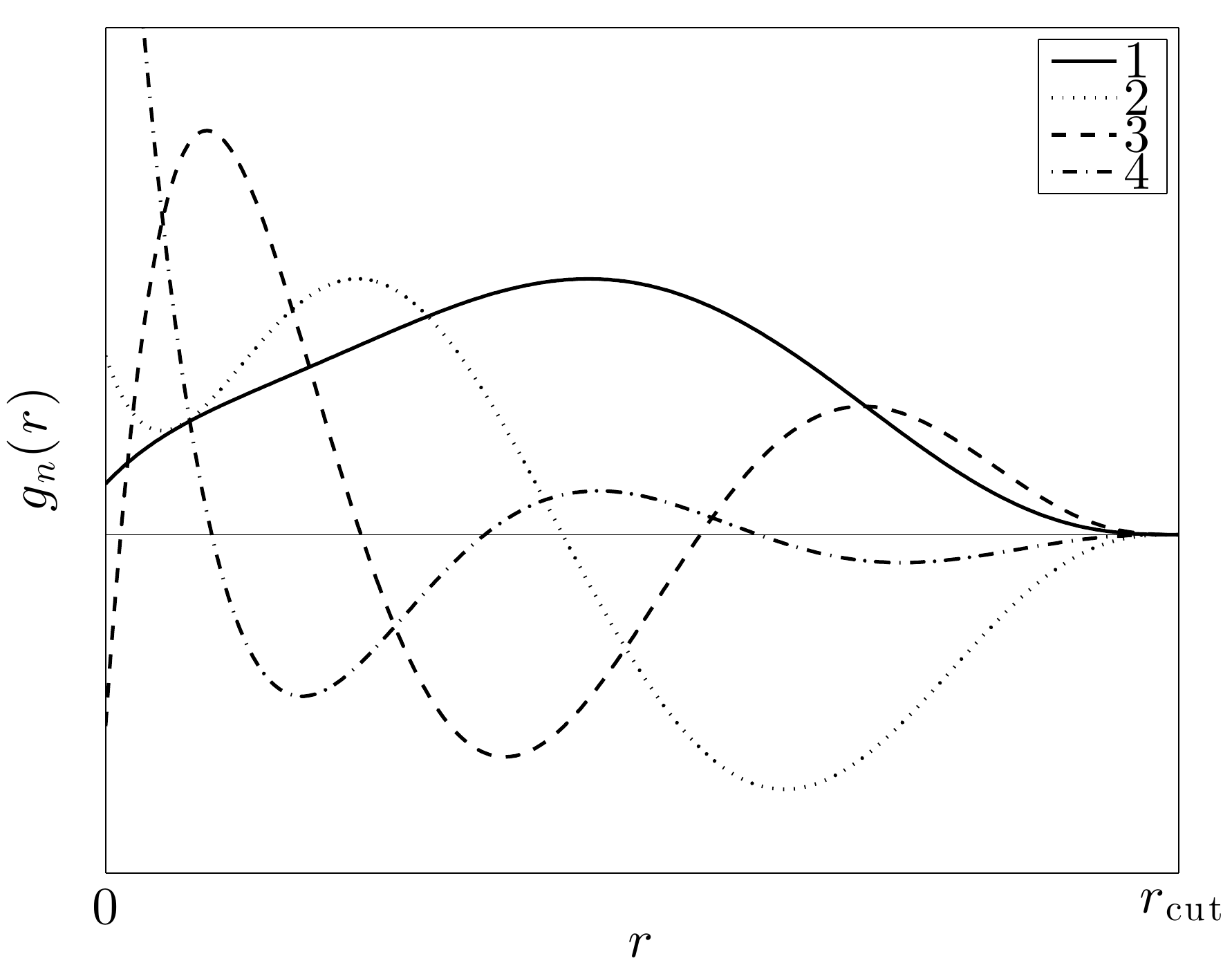}
\caption{Example of radial basis functions $g_n(r)$, as defined in equation~\refpar{eq:afs_radial} for $n=1,2,3,4$.}
\label{fgr:radial}
\end{figure}

Another way to avoid radial decoupling is to define the rotational invariants in such a way that they couple different radial channels explicitly, for example, as
\begin{align}
&p_{n_1 n_2 l}=\sum_{m=-l}^l c_{n_1 lm}^* c_{n_2 l m} \textrm{ , and}\label{eq:pnnl}\\
&b_{n_1 n_2 l\, l_1 l_2} = \sum_{m=-l}^l \sum_{m_1=-l_1}^{l_1} \sum_{m_2=-l_2}^{l_2} c_{n_1 lm}^*
C_{m m_1 m_2}^{l\,l_1 l_2} \times \nonumber\\
& \times c_{n_2 l_1 m_1} c_{n_2 l_2 m_2} \textrm{.}\nonumber
\end{align}
Here, each invariant has contributions from two different radial basis channels, and so we ensure that they cannot become decoupled, but at the price of increasing the
number of invariants quadratically or even cubically in the number of radial basis functions used.


\subsection{4-dimensional power spectrum and bispectrum}\label{sec:so4}

We now present an alternative to the SO(3) power spectrum and bispectrum  that does not need the explicit introduction of a radial basis set, but still represents atomic neighbourhoods in three-dimensional space.
We start by projecting the atomic neighbourhood density within a cutoff radius $r_\textrm{cut}$ onto the surface of the four-dimensional sphere $S^3$ with radius $r_0$. The surface of $S^3$ is defined as the set of points $\mathbf{s} \in \mathbb{R}^4$, where $s_1^2+s_2^2+s_3^2+s_4^2=r_0^2$, while the polar angles $\phi$, $\theta$ and $\theta_0$ of $\mathbf{s}$ are defined so that
\begin{subequations}
\begin{align*}
&s_1=r_0 \cos \theta_0 \\
&s_2=r_0 \sin \theta_0 \cos \theta \\
&s_3=r_0 \sin \theta_0 \sin \theta \cos \phi \\
&s_4=r_0 \sin \theta_0 \sin \theta \sin \phi
\textrm{.}
\end{align*}
\end{subequations}
We choose to use the projection
\begin{equation*}
\mathbf{r}\equiv\left(\begin{matrix}x\\y\\z\end{matrix}\right) \rightarrow
\begin{aligned}
&\phi = \arctan(y/x) \\ &\theta= \arccos(z/|\mathbf{r}|) \\ &\theta_0 =\pi |\mathbf{r}|/r_0
\end{aligned}
\textrm{,}
\end{equation*}
where  $r_0 > r_\textrm{cut}$ is a parameter, thus rotations in three-dimensional space correspond to a subset of rotations in  four-dimensional space. This projection is somewhat similar to a Riemann projection, except in that case  $\theta_0$ would be defined as 
\begin{equation*}
\theta_0=2 \arctan (|\mathbf{r}|/2r_0),
\end{equation*}
implying
\begin{equation*}
\theta_0 \approx |\mathbf{r}|/r_0 \text{ for } |\mathbf{r}| \ll r_0 \text{.}
\end{equation*}
In contrast to the Riemann projection, our choice of $\theta_0$ allows more sensitive representation of the entire radial range. The limit $r_0=r_\textrm{cut}$ projects each atom at the cutoff distance to the South pole of the \mbox{4-dimensional} sphere, thus losing all angular information. Too large an $r_0$ would project all positions onto a small surface area of the sphere around the North pole, requiring a large number of basis functions to represent the atomic environment. In practice, a large range of $r_0$ values works well, in particular, we used $r_0 = \tfrac{4}{3}\:r_\textrm{cut}$.

\begin{figure}[htb]
\centering
\includegraphics[width=4cm]{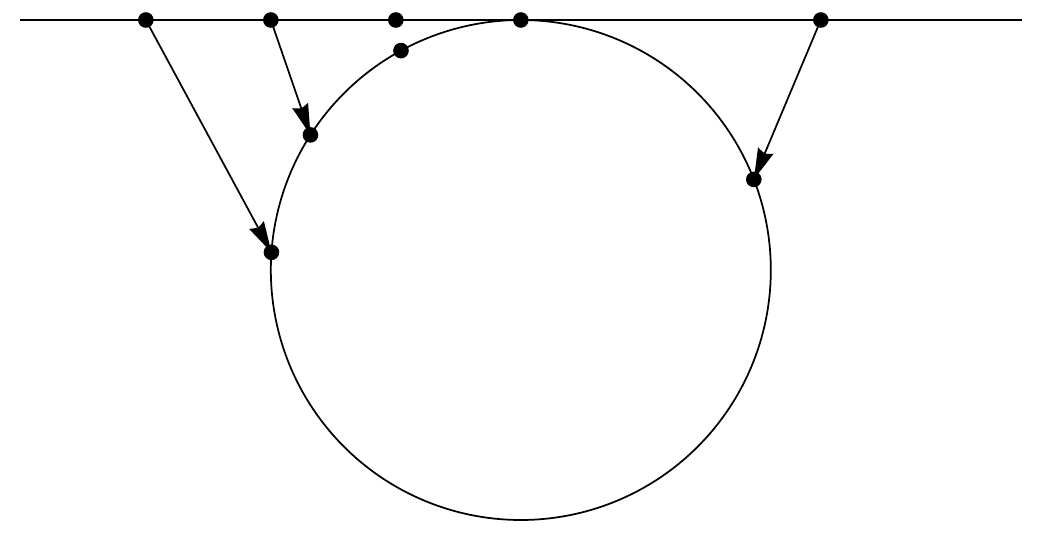}
\includegraphics[width=4cm]{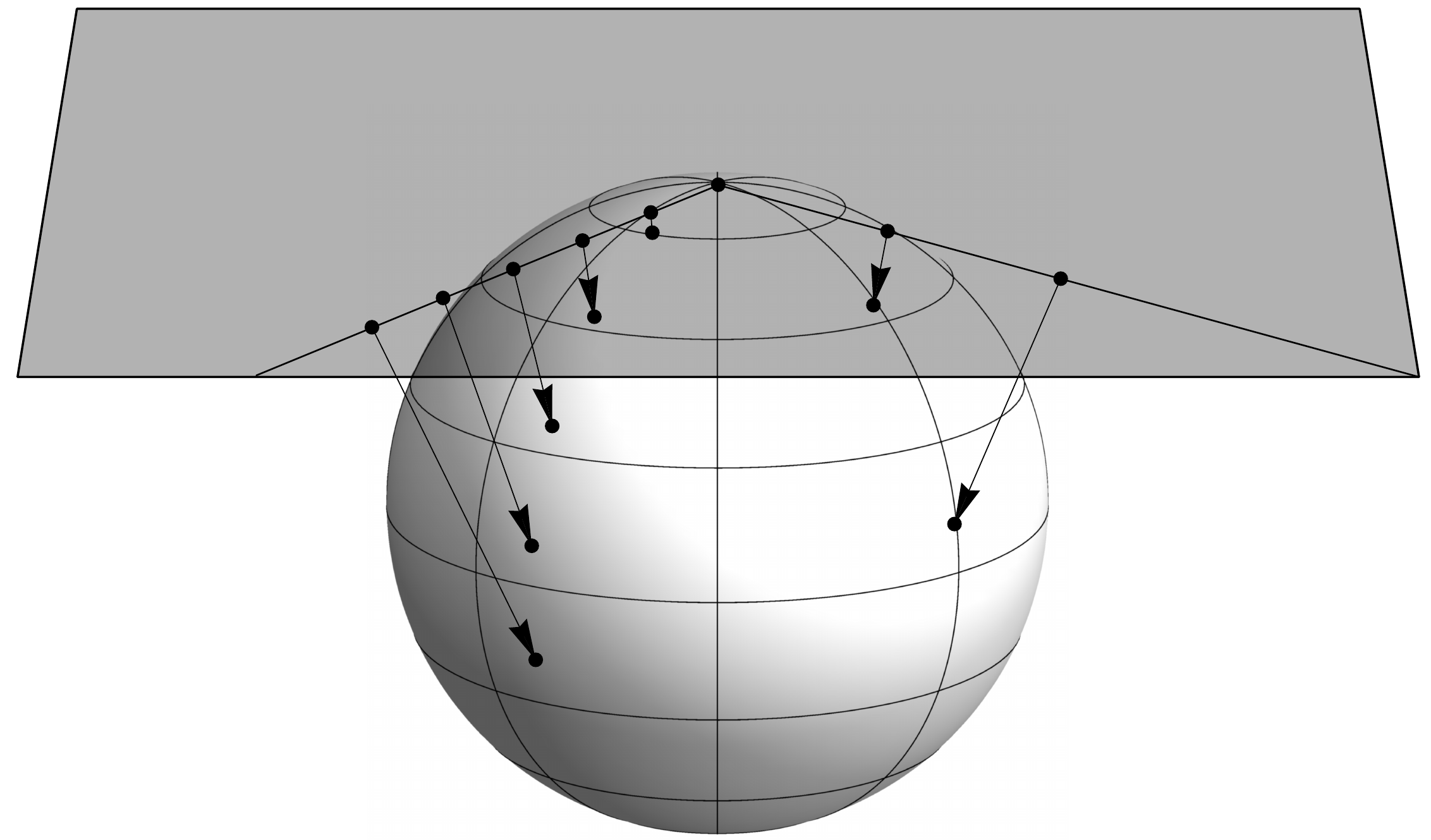}
\caption{Two- and three-dimensional Riemann constructions that map a flat space onto the surface of a sphere in one higher dimension.}
\label{fig:riemann}
\end{figure}

To illustrate the procedure, Figure~\ref{fig:riemann} shows the Riemann projections for one and two dimensions, which can be easily drawn.

An arbitrary function $\rho$ defined on the surface of a 4D sphere can be numerically represented using the hyperspherical harmonic functions $U^j_{m'm}(\phi, \theta, \theta_0)$\cite{Meremianin:2006gs,Varshalovich:1987ul}:
\begin{equation}\label{eq:hyper_expansion}
\rho = \sum _{j=0} ^\infty \sum _{m,m'=-j}^{j} c^j_{m'm} U^j_{m'm} \textrm{,}
\end{equation}
which, in fact, correspond to individual matrix components of the Wigner (\ie rotational) matrices, as defined in equation~\refpar{eq:wigner}. In this case the arguments represent a rotation by $\theta_0$ around the vector pointing in the $(\phi,\theta)$ direction, which can be
transformed to the conventional Euler-angles, and $j$ takes half-integer values.

The hyperspherical harmonics form an orthonormal basis for $L_2(S^3)$, thus the expansion coefficients
$c^j_{m'm}$ can be calculated via
\begin{equation*}
c^j_{m'm} = \langle U^j_{m'm} | \rho \rangle \textrm{,}
\end{equation*}
where $\langle . | . \rangle$ denotes the inner product defined on the four-dimensional hypersphere:
\begin{multline*}
\langle f | g \rangle = \int_0^\pi \mathrm{d} \theta_0 \: \sin^2 \theta_0 \int_0^\pi \mathrm{d} \theta \: \sin \theta \int_0^{2\pi} \mathrm{d} \phi \\ f^*(\theta_0,\theta,\phi) \: g(\theta_0,\theta,\phi)
\textrm{.}
\end{multline*}
Although the coefficients $c^j_{m'm}$ have two indices besides $j$, for each $j$ it is convenient to collect them into a single vector $\mathbf{c}^j$. Similarly to the three-dimensional case, rotations act on the hyperspherical harmonic functions as
\begin{equation*}
\hat{R} U^j_{m_1' m_1} = \sum_{m_2' m_2} R^j_{m_1' m_1 m_2' m_2} U^j_{m_2' m_2} \textrm{,}
\end{equation*}
where the matrix elements $R^j_{m_1' m_1 m_2' m_2}$ are given by
\begin{equation*}
R^j_{m_1' m_1 m_2' m_2} = \langle U^j_{m_1' m_1} | \hat{R} | U^j_{m_2' m_2} \rangle \textrm{.}
\end{equation*}
Hence the rotation $\hat{R}$ acting on $\rho$ transforms the coefficient vectors $\mathbf{c}^j$
according to
\begin{equation*}
\mathbf{c}^{j} \to \mathbf{R}^j \mathbf{c}^j \textrm{.}
\end{equation*}
The unitary $\mathbf{R}^j$ matrices are the SO(4) analogues of the Wigner-matrices $\mathbf{D}^l$ of the SO(3) case above, and it can be shown that the direct product of the four-dimensional rotation matrices decomposes according to
\begin{equation*}
\mathbf{R}^{j_l} \otimes \mathbf{R}^{j_2} = \left( \mathbf{H}^{j_1 j_2} \right)^\dagger
\left[ \bigoplus_{j=|j_1-j_2|}^{j_1+j_2} \mathbf{R}^{j} \right] \mathbf{H}^{j_1 j_2} \textrm{,}
\end{equation*}
which is the 4-dimensional analogue of equation~\refpar{eq:cross_wigner_compact}.
The coupling constants $\mathbf{H}^{j_1 j_2}$, or Clebsch-Gordan coefficients of SO(4) are\cite{Meremianin:2006gs,Caprio:2010ho}
\begin{equation*}
H^{j m m'}_{j_1 m_1 m_1',j_2 m_2 m_2'} \equiv
C_{m m_1 m_2}^{j\, j_1 j_2} C_{m' m'_1 m'_2}^{j\,j_1 j_2}
\textrm{.}
\end{equation*}
The rest of the derivation continues analogously to the 3D case, and finally we arrive at the expression for the SO(4) bispectrum elements
\begin{align*}
B_{j\, j_1 j_2} &= \sum_{m'_1, m_1 = -j_1}^{j_1} c^{j_1}_{m'_1 m_1} \sum_{m'_2, m_2 = -j_2}^{j_2} c^{j_2}_{m'_2 m_2} \times \\
& \times \sum_{m', m = -j}^{j} C^{j\,j_1 j_2}_{m m_1 m_2} C^{j\,j_1 j_2}_{m' m'_1 m'_2} \bigl( c^j_{m'm} \bigr)^*
\textrm{,}
\end{align*}
while the SO(4) power spectrum is
\begin{equation*}
P_j = \sum_{m', m = -j}^{j} \left( c^j_{m'm} \right)^* c^j_{m'm} \textrm{.}
\end{equation*}

The SO(4) bispectrum is invariant to rotations of four-dimensional space, which
include three-dimensional rotations. However, there are
additional rotations, associated with the third polar angle $\theta_0$, which, in our case,
represents the radial information. In order to eliminate the unphysical invariance with respect to rotations along the third polar angle, we modify the atomic density as
\begin{equation*}
\rho (\mathbf{r}) = \delta(\mathbf{0}) + \sum_i \delta(\mathbf{r}-\mathbf{r}_{i}) \textrm{,}
\end{equation*}
\ie we add the central atom --- with the coordinates $(0,0,0)$ --- as a fixed reference point, anchoring the neighbourhood. The resulting invariants $B_{j\,j_1 j_2}$ have only three indices, but contain both radial and angular information, and have the required symmetry properties. There are no adjustable parameters in the definition of these invariants, apart from the projection parameter $r_0$ discussed above.

The number of components in the truncated representation depends on the band limit $j_{\mathrm{max}}$ in the expansion~\refpar{eq:hyper_expansion}.
For symmetry reasons, the bispectrum components with non-integer $j_1+j_2+j$ change sign under reflection and, because of this reason, we omitted them. Just as in the 3D case, the representation is probably over-complete, \ie most of the bispectrum components are redundant. To reduce the number of redundant elements, we only used the `diagonal' components, \ie $j_1=j_2$. Table~\ref{tbl:num_bisp} shows the number of bispectrum elements for increasing band limit values.

\begin{table}[h]
\small
\caption{Number of  components in the full and diagonal bispectrum as a function of the band limit $j_\textrm{max}$.}
\label{tbl:num_bisp}
\begin{ruledtabular}
  \begin{tabular}{ccccccccccc}
    $j_{\mathrm{max}}$ & 0 & $\tfrac{1}{2}$ & 1 & $\tfrac{3}{2}$ & 2 & $\tfrac{5}{2}$ & 3 & $\tfrac{7}{2}$ & 4 & $\tfrac{9}{2} $ \B \\ 
    \colrule
    $B_{j\, j_1 j_1}$  & 1 & 2                  & 5 & 7                 & 12 &15               & 22 &   26            & 35 &  40             \\ 
    $B_{j\, j_1 j_2}$  & 1 & 4                  & 11 & 23            & 42  & 69             & 106 & 154        & 215 & 290           \\ 
  \end{tabular}
\end{ruledtabular}
\end{table}


\subsection{Parrinello-Behler descriptor}\label{subsection:pb}

We include in the tests below  the descriptor suggested by Parrinello and Behler\cite{Behler:2007th} using the parameters published recently \cite{Artrith:2012fw} (and henceforth termed PB). The two- and three-body symmetry functions (in their terminology) are,
\begin{equation*}
G^2_{\alpha} = \sum_{i} = \exp \left[ -\eta_\alpha (r_{i}-R_{s\alpha})^2 \right] f_c(r_{i})
\end{equation*}
and
\begin{multline*}
G^4_{\alpha} = 2^{1-\zeta_\alpha} \sum_{i,i^\prime} ( 1+ \lambda_\alpha \cos \theta_{ii^\prime} )^{\zeta_\alpha} \times \\
\exp \left[ -\eta_\alpha (r_{i}^2 + r_{i^\prime}^2 + r_{ii^\prime}^2) \right] f_c(r_{i}) f_c(r_{i^\prime}) f_c(r_{ii^\prime})
\textrm{,}
\end{multline*}
where the cutoff function is defined as
\begin{equation*}
f_c(r) = \left\{
\begin{array}{c l}
\left[ \cos \left( \frac{\pi r}{r_\textrm{cut}} \right) + 1 \right]/2 & \quad \text{for } r \le r_\textrm{cut} \\
0 & \quad \text{for } r > r_\textrm{cut}
\end{array}
\right.
\textrm{.}
\end{equation*}
Different values of the parameters $\eta, R_s, \zeta, \lambda$ can be used to generate an arbitrary number of invariants.
\begin{figure}[htb]
\includegraphics[width=8.6cm]{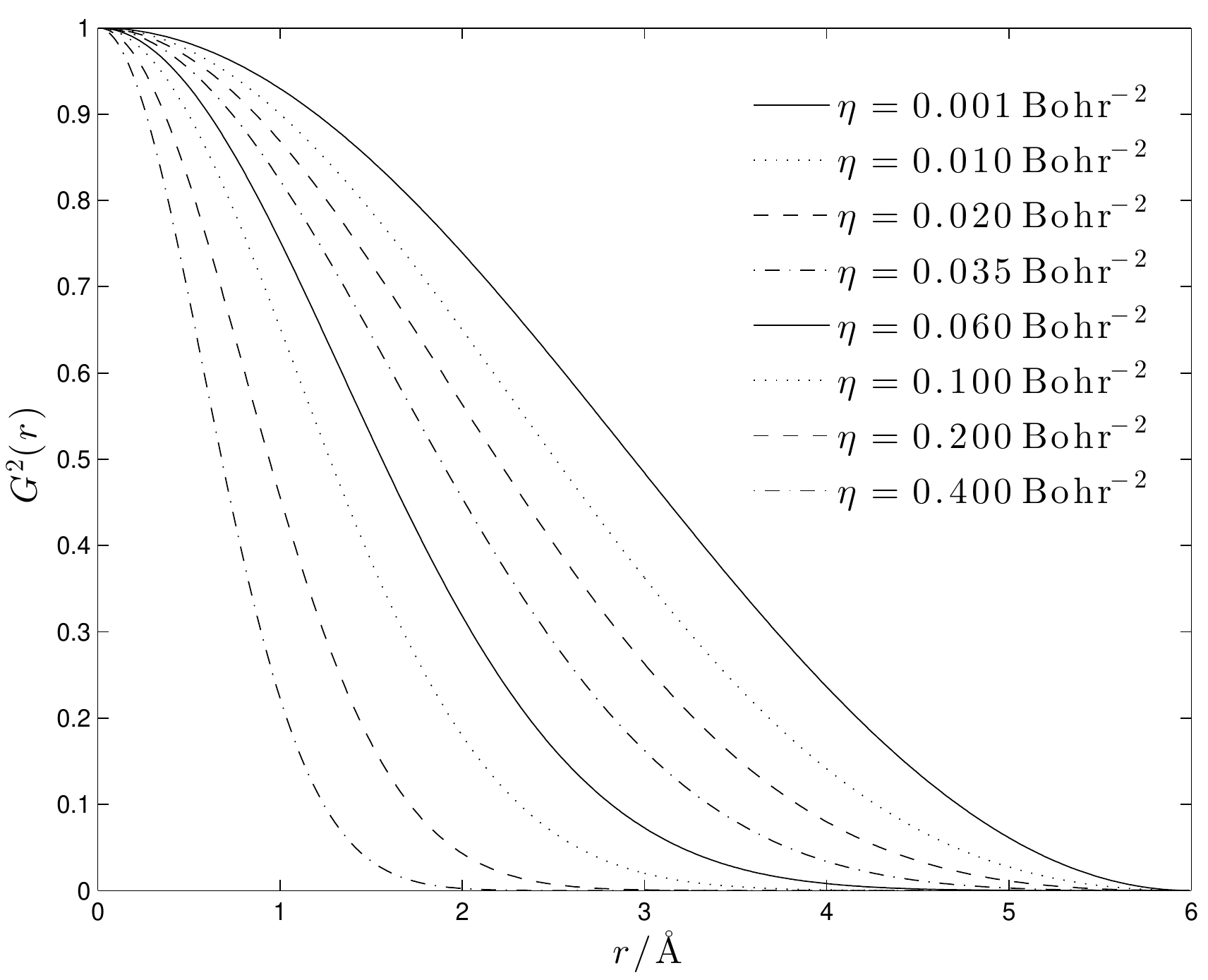}
\caption{Radial basis functions $G^2$ in the Parrinello-Behler (PB) type descriptors\cite{Artrith:2012fw}.}
\end{figure}

\subsection{Angular Fourier Series}
Notice that the angular part of the power spectrum, bispectrum (section~\ref{sec:bispectrum}) and the descriptors defined by Parrinello and Behler (section \ref{subsection:pb}) are all simple polynomials of the canonical set $\sum_{ii^\prime} \cos^m \theta_{ii^\prime}$ for integer $m$, which, in turn, are sums of powers of the basic invariants of Weyl. We include in the tests in the next section a further descriptor set, which we call the Angular Fourier Series (AFS) descriptor, formed by a system of orthogonal polynomials of the basic invariants, conveniently chosen as the Chebyshev-polynomials $T_l(x)$, as
\begin{equation*}
T_l(\cos \theta) = \cos (l \theta)
\textrm{,}
\end{equation*}
and incorporate the radial information using the basis functions defined in equation~\refpar{eq:afs_radial}, leading to
\begin{equation*}
\mathrm{AFS}_{n,l}=\sum_{i,i^\prime} g_n(r_{i}) g_n(r_{i^\prime}) \cos(l \theta_{ii^\prime})
\textrm{.}
\end{equation*}
\begin{figure}[htb]
\includegraphics[width=8.6cm]{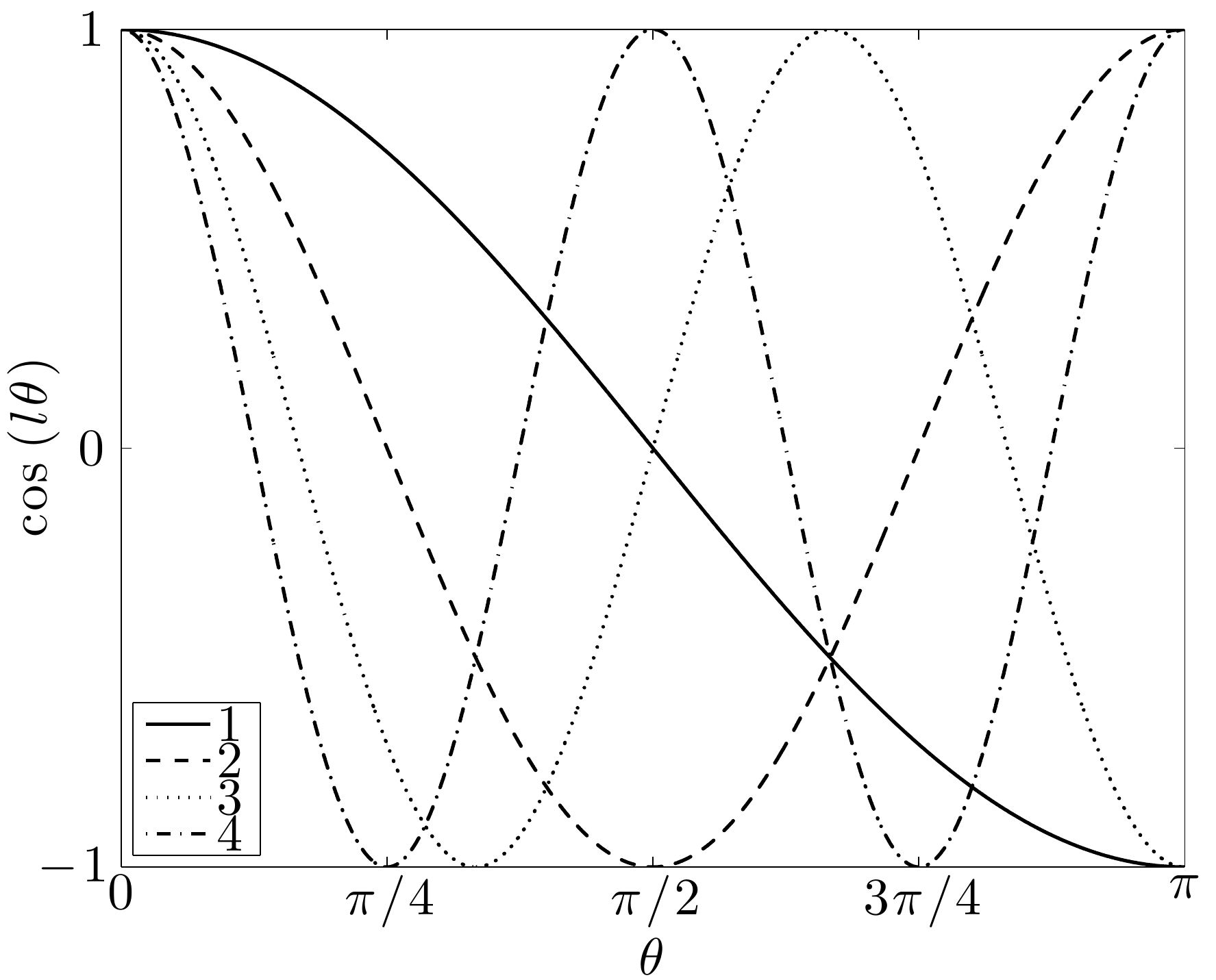}
\caption{Examples of the angular basis functions for $l_\mathrm{max}=4$ of the AFS descriptor.}
\label{fgr:angular}
\end{figure}


\section{A similarity measure between atomic environments}\label{sec:soap}
It is clear from the preceding section that there is a lot of freedom in constructing descriptors, e.g. in the choice of angular band limit, the radial basis and also which subset of the basis elements are actually used. As we have shown in section \ref{sec:pes}, the key to PES fitting are not the descriptors \perse, but the similarity measure $K(\mathbf{q}, \mathbf{q}^\prime)$ that is constructed from the descriptors.
This suggests an alternative approach, in which descriptors are bypassed altogether, and a similarity measure between atomic neighbourhoods is constructed directly. The criteria for a good similarity measure is not only that it be invariant to symmetry operations of the atoms of each environment and have a well-defined limit when comparing two identical or two very different environments, but also that the it change smoothly with the Cartesian atomic coordinates. 

We define the similarity of two atomic environments directly as the inner product of two atomic neighbour densities $\rho$ and $\rho^\prime$ (defined in equation \refpar{eq:atomic_density}), as the overlap
\begin{equation}\label{eq:similarity_rhoi_rhoj}
S(\rho,\rho^\prime) = \int \rho(\mathbf{r}) \rho^\prime(\mathbf{r}) \diffd \mathbf{r}  \textrm{.}
\end{equation}
This clearly satisfies the permutational invariance criterion. Integrating equation \refpar{eq:similarity_rhoi_rhoj} over all possible rotations of one of the environments leads to a rotationally invariant similarity kernel
\begin{multline}\label{eq:simkernelrho}
k(\rho,\rho^\prime) = \int \left|S(\rho,\hatR \rho^\prime) \right|^n \diffd \hatR = \\
=\int \diffd \hatR \left| \int \rho(\mathbf{r}) \rho^\prime(\hatR \mathbf{r}) \diffd \mathbf{r} \right|^n
\textrm{,}
\end{multline}
It is easy to see that for $n=1$, all angular information -- the relative orientation of individual atoms -- is lost because the order of the two integrations can be exchanged, but for $n \ge 2$ the kernel retains the angular information of the original environments. The obvious practical difficulty with this construction is the evaluation of the angular integral, which is addressed next.  

\subsection{Analytic evaluation a smooth similarity kernel}
Retaining the Dirac-delta functions in the definition of the atomic neighbour density would lead to a discontinuous similarity kernel in that the dissimilarity between two environments with very close but not identical atomic positions would be large. Therefore, instead of the Dirac-delta functions, we construct the atomic neighbour density using Gaussians,  expanded in terms of spherical harmonic functions as \cite{Kaufmann:1989tq}
\begin{multline}
\exp \left(-\alpha | \mathbf{r} - \mathbf{r}_{i} |^2 \right) = \\
4\pi \exp \left[-\alpha (r^2+r^2_{i}) \right] \sum_{lm} \iota_l( 2 \alpha r r_{i} ) \,
Y_{lm}(\hatbfr) \, Y^*_{lm}(\hatbfr_{i})
\textrm{,}
\end{multline}
where $\iota_l$ are the modified spherical Bessel functions of the first kind. The atomic neighbour density function is then defined as a sum of Gaussians with one centred on each neighbour,
\begin{equation}
\label{eq:expexpansion}
\rho(\mathbf{r}) = \sum_i \exp \left(-\alpha | \mathbf{r} - \mathbf{r}_{i} |^2 \right) = 
\sum_i \sum_{lm} c^{i}_{lm} (r) Y_{lm}(\hatbfr)
\textrm{,}
\end{equation}
where
\begin{equation*}
c^{i}_{lm} (r) \equiv 4\pi \exp \left[-\alpha (r^2+r^2_{i}) \right] \iota_l( 2 \alpha r r_{i} ) Y^*_{lm}(\hatbfr_{i})
\textrm{.}
\end{equation*}

The overlap between an atomic environment (unprimed) and a rotated environment (primed) is
\begin{widetext}
\begin{multline*}
S(\hatR) \equiv S(\rho,\hatR \rho^\prime)  = \int \diffd \mathbf{r} \, \rho(\mathbf{r}) \rho^\prime(\hatR \mathbf{r}) =
\sum_{i, {i^\prime}} \sum_{
\mbox{\scriptsize
$\begin{array}{c}
l,m \\
l',m',m''
\end{array}
$}}
D^{l'}_{m'm''} (\hatR) \int \diffd r \, c^{i*}_{lm}(r) c^{{i^\prime}}_{l'm'}(r) 
\int \diffd \hatbfr \: Y_{lm}^* (\hatbfr) Y_{l'm''} (\hatbfr) = \\
= \sum_{i,{i^\prime}} \sum_{l,m,m'} \tilde I^l_{mm'}(\alpha,r_{i},r_{i^\prime}) D^{l}_{mm'} (\hatR) = 
\sum_{l,m,m'} I^l_{mm'} D^{l}_{mm'}(\hatR)
\textrm{,}
\end{multline*}
\end{widetext}
where the integral of the coefficients is
\begin{multline*}
\tilde I^l_{mm'}(\alpha,r_{i},r_{i^\prime}) = \\
4\pi \exp \left( - \alpha (r^2_{i} + r^2_{i^\prime})/2 \right) 
\iota_l \left( \alpha r_{i} r_{i^\prime } \right) Y_{lm} (\hatbfr_{i}) Y^*_{lm}(\hatbfr_{i^\prime})
\textrm{,}
\end{multline*}
and
\begin{equation}
\label{eq:integral_ilmm}
I^l_{mm'} \equiv \sum_{i, i^\prime} \tilde I^l_{mm'}(\alpha,r_{i},r_{i^\prime })
\textrm{.}
\end{equation}

The rotationally invariant  kernel with $n=2$ then becomes
\begin{multline}
\label{eq:kernel_ilmm}
k(\rho,\rho^\prime) = \int \diffd \hatR \, S^*(\hatR) S(\hatR) = \\
=\sum_{
\mbox{\scriptsize
$\begin{array}{l}
l,m,m' \\
\lambda,\mu,\mu'
\end{array}
$}}
\left( I^l_{mm'} \right)^* I^\lambda_{\mu \mu'} 
\int \diffd \hatR \, D^*(\hatR)^{l}_{mm'} D(\hatR)^{\lambda}_{\mu \mu'} = \\
=\sum_{l,m,m'} \left( I^l_{mm'} \right)^*  I^l_{mm'}
\textrm{,}
\end{multline}
where we used the orthogonality of the Wigner-matrices. Although in practice we always use $n=2$, it is easy to derive the kernel for any arbitrary order $n$ using the Clebsch-Gordan series in equation \refpar{eq:CG_series}.
For $n=3$, using the fact that $S$, as defined in equation \refpar{eq:similarity_rhoi_rhoj} is real and positive,
\begin{equation*}
k(\rho,\rho^\prime) = \int \diffd \hatR \; S(\hatR)^3
\textrm{,}
\end{equation*}
which can be shown to be
\begin{equation}
\label{eq:kernel_third}
k(\rho,\rho^\prime) = \sum I^{l_1}_{m_1m_1'} I^{l_2}_{m_2m_2'} I^{l}_{mm'} C^{lm}_{l_1 m_1 l_2 m_2} C^{lm'}_{l_1 m_1' l_2 m_2'}
\textrm{.}
\end{equation}
Raising a positive definite function to a positive integer power yields a function that is similarly positive definite.
In our context, raising $k$ to some power $\zeta\geq 2$ has the effect of accentuating the sensitivity of the kernel to changing the atomic positions, which we generally found to be advantageous in experiments.
Therefore, following normalization by dividing by $\sqrt{k(\rho,\rho)k(\rho',\rho')}$, we define the general form of what we call the Smooth Overlap of Atomic Positions (SOAP) kernel as
\begin{equation}\label{eq:similarity_kernel}
K(\rho,\rho^\prime) = \Biggl( \frac{k(\rho,\rho^\prime)}{\sqrt{ k(\rho,\rho) k(\rho^\prime,\rho^\prime) }} \Biggr)^\zeta
\textrm{,}
\end{equation}
where $\zeta$ is any positive integer.

\subsection{Radial basis and relation to spectra}
\begin{figure}[h]
  \includegraphics[width=8.6cm]{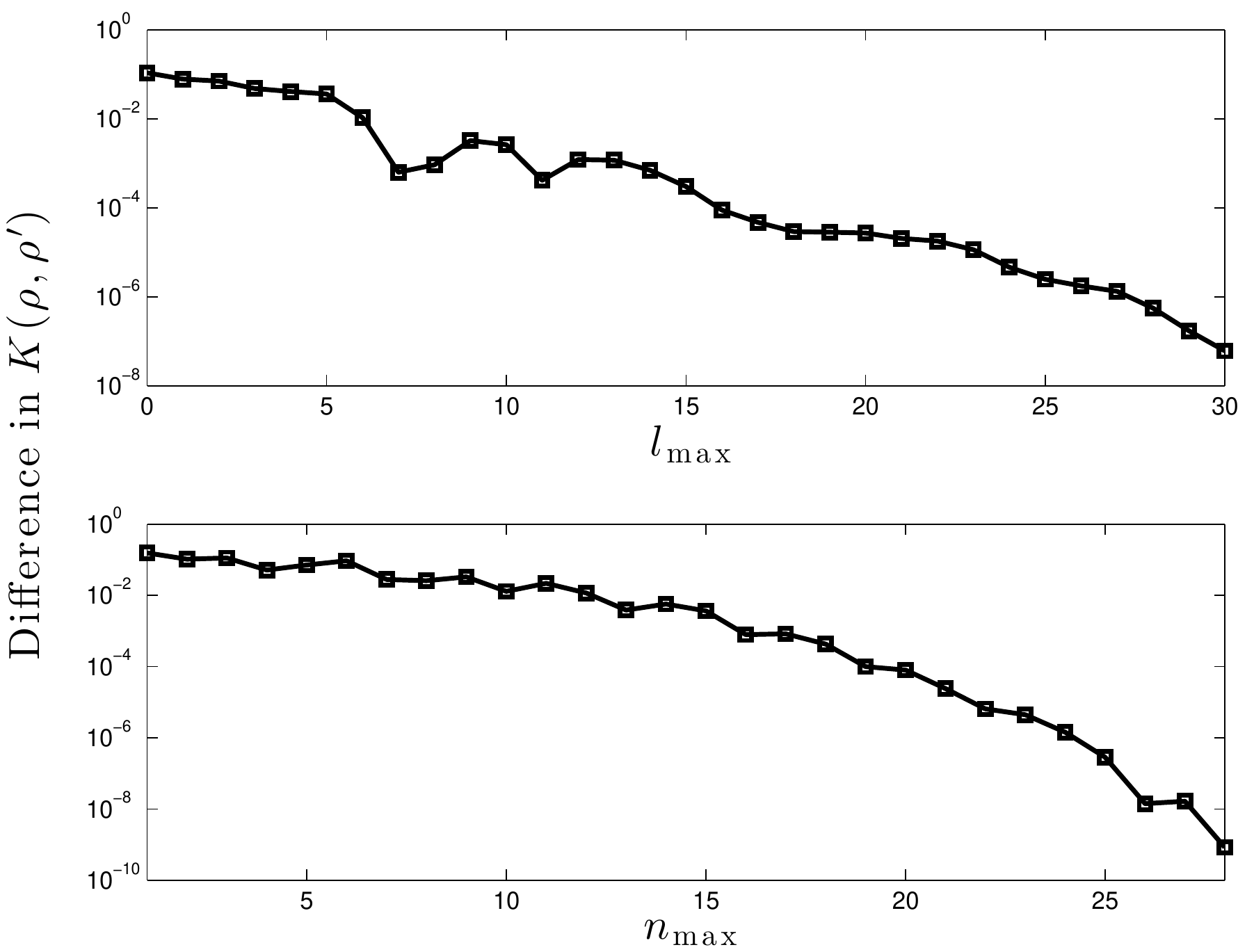}
  \caption{Convergence of the similarity kernel $K(\rho,\rho^\prime)$ of two arbitrary atomic environments with 15 neighbours at different sizes of basis expansion. We used the parameters $\alpha=0.4$ and $\zeta=1.0$, and the converged kernel is $K(\rho,\rho^\prime)=-0.842735$. The top panel shows the convergence of the kernel, evaluated according to equation~\refpar{eq:kernel_ilmm}, with increasing number of spherical harmonics functions. The bottom panel shows the convergence of the kernel, evaluated according to equation~\refpar{eq:kernel_power}, with increasing number of radial basis functions, while keeping $l_\textrm{max}=16$. }
  \label{fgr:soap_convergence}
\end{figure}

Note that $I^l_{mm^\prime}$ needs to be computed for each pair of neighbours, which can become expensive for a large number of neighbours.
If we expand equation~\refpar{eq:expexpansion} using radial basis functions $g_n(r)$,  the atomic neighbour density function becomes
\begin{equation}
\label{eq:soap_radialexp}
\rho(\mathbf{r}) = \sum_i \exp \left(-\alpha | \mathbf{r} - \mathbf{r}_{i} |^2 \right) = 
\sum_{nlm} c_{nlm} g_n(r) Y_{lm}(\hatbfr)
\textrm{,}
\end{equation}
and similarly, the $\rho^\prime$ environment is expanded using coefficients $c^\prime_{nlm}$.
If the radial basis functions form an orthonormal basis, \ie
\begin{equation*}
\int \diffd r g_n(r) g_{n^\prime}(r) = \delta_{nn^\prime}
\textrm{,}
\end{equation*}
the sum in equation \refpar{eq:integral_ilmm} becomes
\begin{equation}
\label{eq:integral_rad}
I^l_{mm^\prime} = \sum_n c_{nlm} \left(c^\prime_{nlm^\prime}\right)^*
\textrm{.}
\end{equation}
The significance of this result becomes apparent when substituting equation \refpar{eq:integral_rad} into equation \refpar{eq:kernel_ilmm} to obtain
\begin{multline}
\label{eq:kernel_power}
k(\rho,\rho^\prime) = \sum_{n,n',l,m,m'}  c_{nlm} (c^\prime_{nlm'})^* (c_{nlm})^* c^\prime_{n'lm'}  \\
\equiv \sum_{n,n',l} p_{nn'l} p^\prime_{nn'l}
\textrm{,}
\end{multline}
since
\begin{equation}
p_{nn'l} \equiv \sum_m c_{nlm} (c_{n'lm})^*
\textrm{,}
\end{equation}
is exactly the power spectrum (\cf equation \refpar{eq:pnnl}), and, analogously, $p^\prime_{nn'l}$ is the power spectrum of the primed environment. Furthermore, the kernel in equation \refpar{eq:kernel_power} is the dot-product of the power spectra, \cf equation \refpar{eq:dp-kernel}.
Analogously, the kernel for $n=3$, defined in equation \refpar{eq:kernel_third}, can be expressed as
\begin{equation}
\label{eq:kernel_bispectrum}
k(\rho,\rho^\prime) =
\sum_{
\mbox{\scriptsize
$\begin{array}{c}
n_1,n_2,n \\
l_1,l_2,l
\end{array}
$}} 
b_{n_1 n_2 n l\, l_1 l_2} b^\prime_{n_1 n_2 n l\, l_1 l_2}
\textrm{,}
\end{equation}
where
\begin{equation*}
b_{n_1 n_2 n \; l_1 l_2 l} \equiv \sum c_{n_1 l_1m_1} c_{n_2 l_2 m_2} (c_{nlm})^* C^{lm}_{l_1 m_1 l_2 m_2}
\textrm{,}
\end{equation*}
\cf equation \refpar{eq:bispectrum_so3}, and $\mathbf{b}^\prime$ is analogously the bispectrum of the primed environment.
In Figure~\ref{fgr:soap_convergence} we show the convergence of the similarity kernel \refpar{eq:kernel_power} with increasing number of angular and radial basis functions in the expansion. 

In this section we started out by taking a different approach to the problem of comparing neighbour environments, defining the SOAP similarity kernel \refpar{eq:similarity_kernel} directly, rather than going via a descriptor.
Equations \refpar{eq:kernel_power} and \refpar{eq:kernel_bispectrum}, however, reveal the relation between SOAP and the SO(3) power spectrum and bispectrum: SOAP is equivalent to using the SO(3) power or bispectrum descriptor together with Gaussian atomic neighbour density contributions and a dot product covariance kernel. The advantage of SOAP over the previous descriptors is that it eliminates some of the ad hoc choices that were needed before, while retaining control over the smoothness of the similarity measure using $\alpha$, the width of the Gaussians in defining the atomic neighbourhood density in equation \refpar{eq:expexpansion} and its sensitivity using the exponent $\zeta$ in equation \refpar{eq:similarity_kernel}. 

\section{Numerical results}\label{sec:results}

We have derived methods to transform atomic neighbourhoods to descriptors that are invariant under the required symmetry operators, however, their relative merit for fitting potential energy surfaces remains to be seen. The faithfulness of the representation, \ie that no genuinely different configurations should map onto the same descriptor, is of particular interest. As the inverse transformation from the descriptor to atomic coordinates --- apart from the simplest cases --- is not available, we describe numerical experiments in which we attempt to reconstruct atomic coordinates from descriptors, up to rotations, reflections and permutations. Descriptors which severely fail in this test are unlikely to be good for fitting potential energy surfaces, because entire manifolds of neighbour environments that are genuinely different with widely varying true energies will be assigned the same descriptor, resulting in fitted PES with many degenerate modes. We  compare and test the performance of the various descriptors by generating potential energy surfaces for silicon clusters and the bulk crystal using our Gaussian Approximation Potentials framework\cite{Bartok:2010fj}.

\subsection{Reconstruction experiments}
Recall that the elements $\mathbf{r}_{i} \cdot \mathbf{r}_{i^\prime}$ of $\Sigma$ defined in equation~\refpar{eq:rirj} are an over-complete set of basic invariants, which, in the case of atoms scattered on the surface of a unit sphere ($|\mathbf{r}_{i}|=1$), are the cosines of the bond angles, $\theta_{ii^\prime}$.
Thus, the angular part of all descriptors in section \ref{sec:descriptors} are permutationally invariant functions of the basic invariants in $\Sigma$, and depending on the actual number of descriptor elements used, they may form an incomplete, complete or over-complete representation of the atomic environment. In practice, one would like to use as few descriptors as possible, partly due to computational cost, but also because descriptors that use high exponents of the angles are likely to lead to less smooth PESs, as will be shown below. 

Given $N$ neighbours, the number of independent degrees of freedom in the neighbourhood configuration is $3N-3$, so we need at most this many {\em algebraically independent} descriptor elements. But because the algebraic dependency relationships between the descriptor elements is in general complicated, it is unclear how many  descriptor elements are actually needed in order to make the descriptor complete and thus able to uniquely specify an atomic environment of the $N$ neighbours. However, it is possible to conduct numerical experiments in which we compare the descriptors of a fixed target with that of a candidate structure and minimise the difference with respect to the atomic coordinates of the candidate. In this way we determine if a representation is likely to be complete or not, and in the latter case to characterise the degree of its faithfulness. 

\paragraph*{Descriptor matching procedure.}
The global minimum of the descriptor difference between the target and the candidate is zero and is always attained on a manifold due to the symmetries built into the descriptors, but for an incomplete descriptor, many inequivalent structures will also appear equivalent, thus enlarging the dimensionality of the global minimum manifold. Furthermore, it can be expected that the descriptor difference function has a number of local minima. 

In our experiments we tried to recover a given target structure after randomising its atomic coordinates. For each $n$ ($4 \leq n \leq 19$) we used $10$ different $\textrm{Si}_n$ clusters as targets, obtained from a tight-binding\cite{Porezag:1995tk} molecular dynamics trajectory run at a temperature of  2000~K. For each target cluster, we selected one atom as the origin, randomised the positions of its neighbours by some amount, and then tried to reconstruct the original structure by minimising the magnitude of the difference between the descriptors of the fixed target and the candidate as the atomic positions of the latter were varied. 

In contrast to a general global minimum search problem, we have the advantage of knowing the target value of the objective function at the global minimum. Also, the motivation of our experiments is to find at least one configuration, if it exists, that is genuinely different from the target, but where the descriptors match within a pre-defined numerical tolerance. Thus it is sufficient to perform local, gradient-based optimisations starting from random configurations, and reject all local minima (by noting the small gradient of the objective function while the value of the objective function is not small) until we find one where the objective function --- the difference in the descriptors --- is less than than the specified tolerance. If the configuration thus obtained is genuinely different from the target, the descriptor is shown to be incomplete. 

In order to assess the success of the reconstruction procedure (i.e. whether the target and candidate configurations are genuinely different or not) we employed the reference measures defined in equations~\refpar{eq:d_refp} and \refpar{eq:d_refr}. However, in some cases it was difficult or impossible to find the right rotation $\hat{R}$ in \refpar{eq:d_refr},
whereas $d_\mathrm{ref}$ in \refpar{eq:d_refp} proved reliable. For each $d_\mathrm{ref}$, an initial $\mathbf{P}$ was generated by ordering the atoms according to their distances from the central atom, then the optimal permutation was found using a simple random search in the space of permutations.

\begin{figure}[h]
  \centering
  \includegraphics[width=4cm]{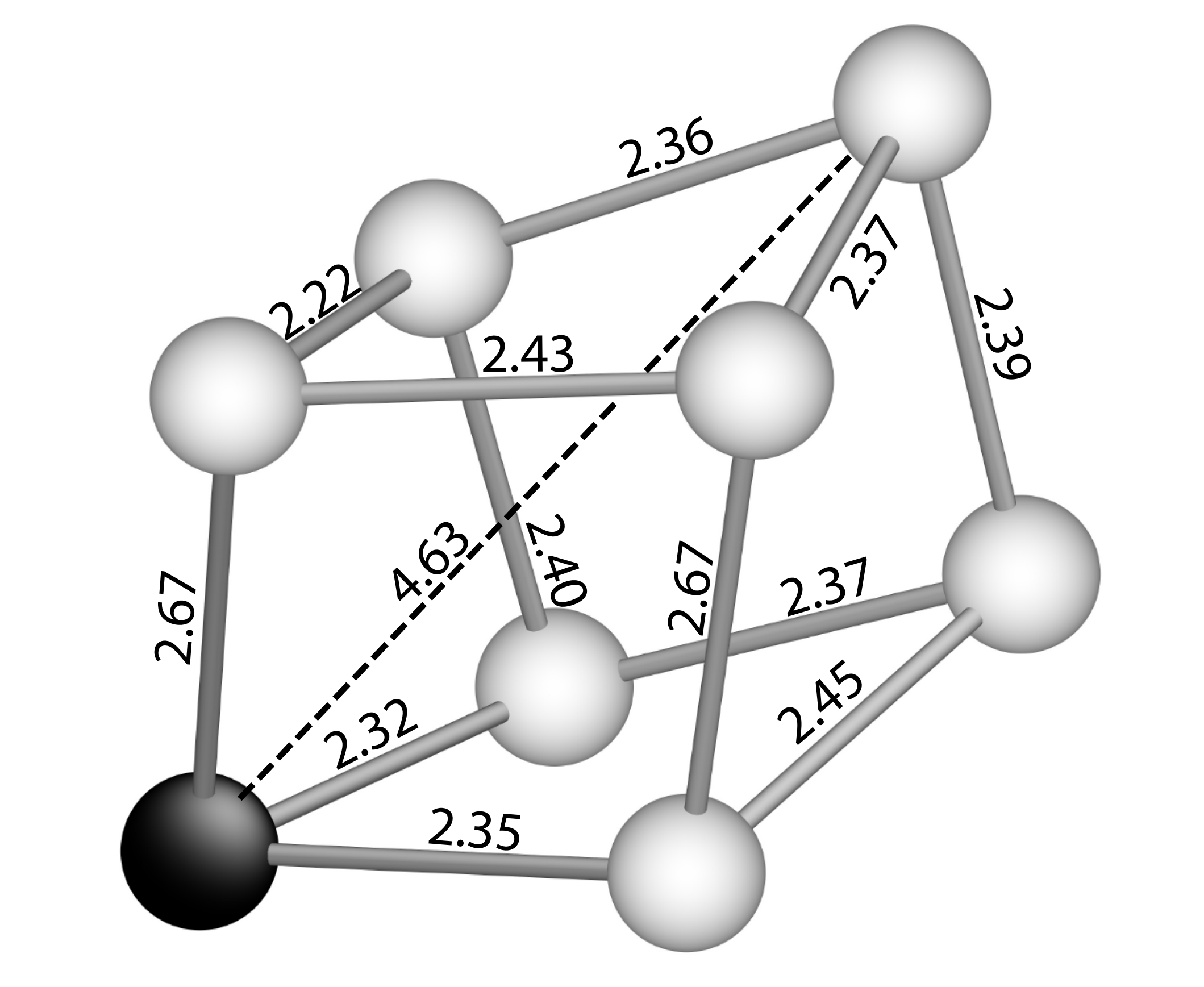}
  \includegraphics[width=4cm]{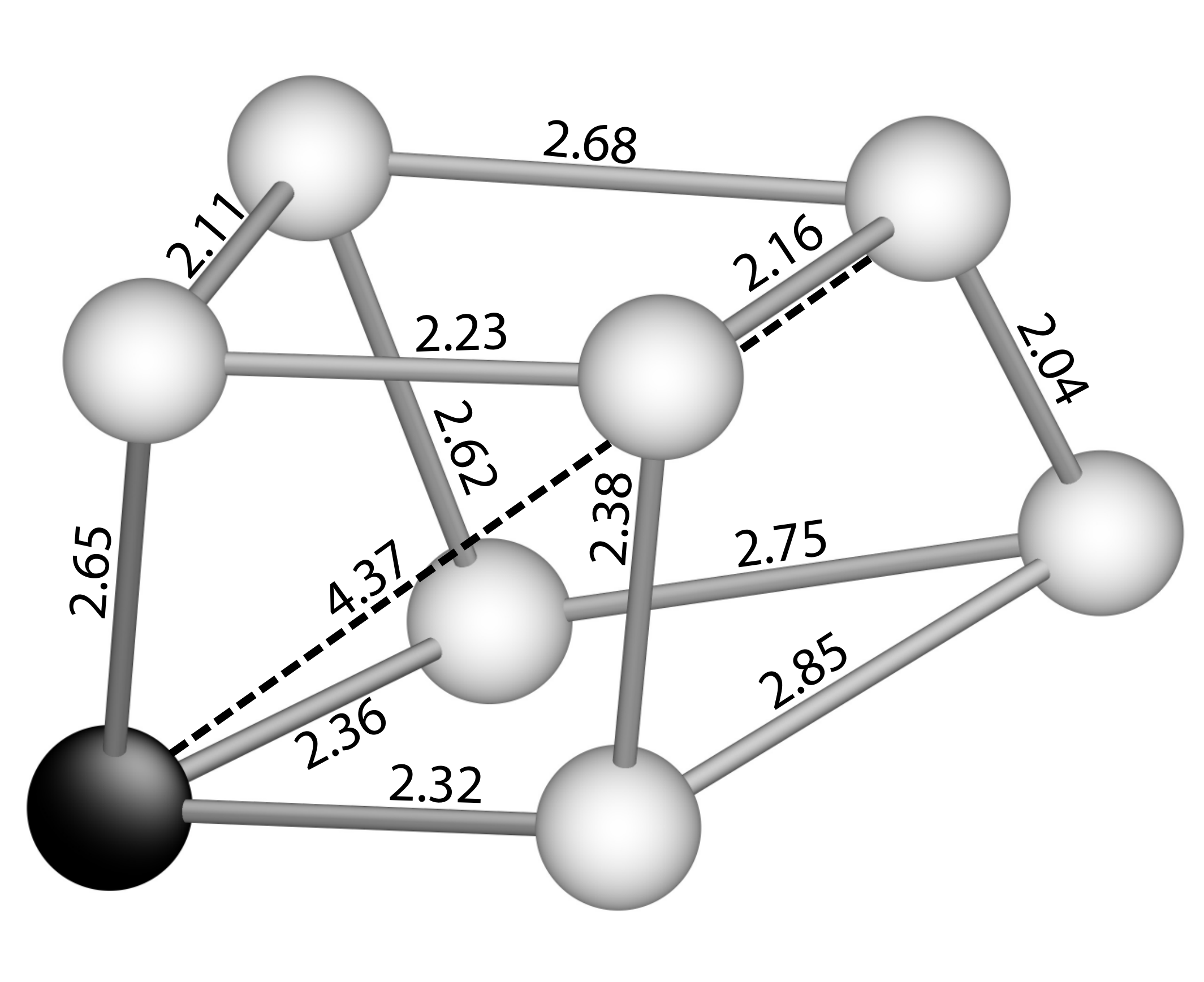}
  \caption{Two $\textrm{Si}_8$ clusters that differ by $d_\textrm{ref} = 4.1 \textrm{\AA}^2$.  The black atoms are taken as the origin in each environment, \ie the centres of rotations. In terms of Parrinello-Behler type descriptors, the difference $\sum_\alpha (G_{\alpha}-G^\prime_{\alpha})^2 $ between the two atomic environments is \hbox{$6\cdot10^{-7}$}. The bond lengths are shown in \AA ngstr\"oms.}
  \label{fgr:dref}
\end{figure}

We minimised the difference between the target and candidate descriptors in the space of atomic coordinates of the latter using the Conjugate Gradients algorithm, stopping the minimisation when either the gradient or the reference distance $d_\textrm{ref}$ became smaller than $10^{-8}$\AA$^2$ and $10^{-2}$\AA$^2$, respectively. In order to ensure that structures deemed non-equivalent by $d_\textrm{ref} > 10^{-2}$\AA$^2$ were genuinely different, we cross-checked them by noting the value of $\Delta$ from equation~\refpar{eq:d_refr} and also employing the atomic fingerprints suggested by Oganov and Valle\cite{Valle:2010jt}. To give a sense of the typical magnitude of the $d_\textrm{ref}$ measure, the actual difference in terms of atomic distances between two example structures is shown in Figure \ref{fgr:dref}. 
 
\begin{figure}[h]
  \includegraphics[width=8.6cm]{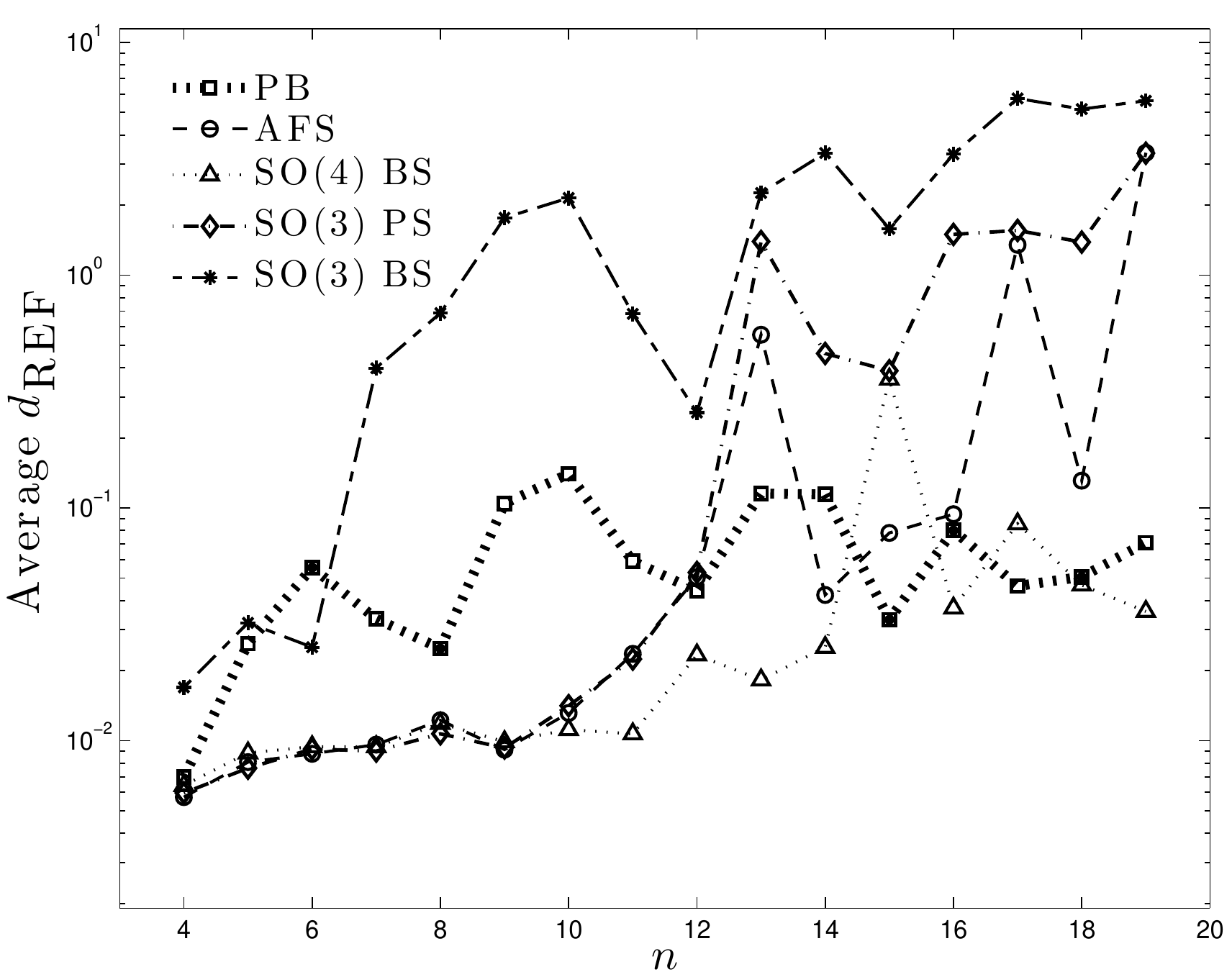}
  \includegraphics[width=8.6cm]{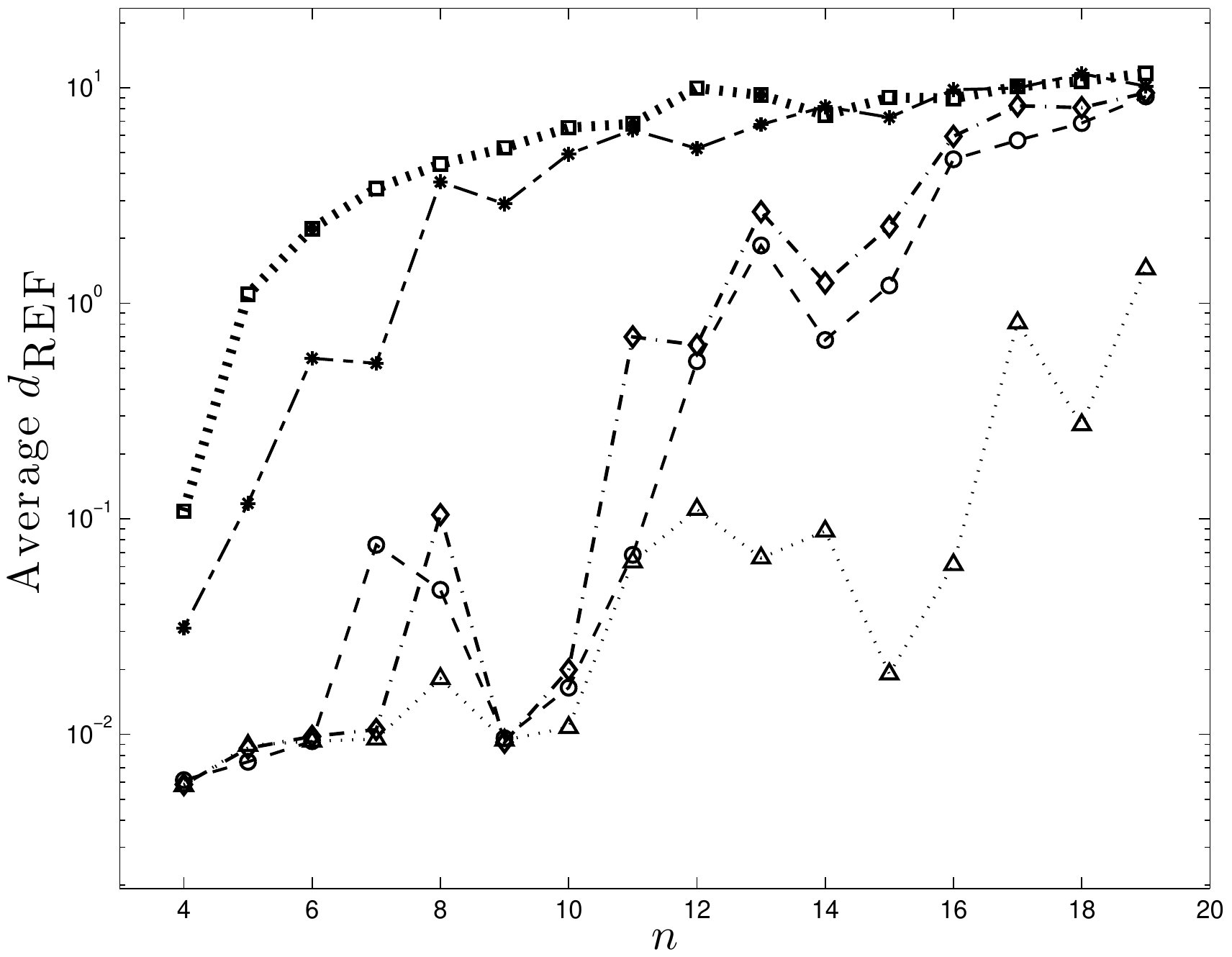}
  \caption{Difference between the target and reconstructed structures after randomisation and minimisation for randomisations of $0.2$ \AA\ (top) and $1.6$ \AA\ (bottom) as a function of the number of atoms in the cluster, $n$, averaged over 10 targets for each cluster size. The radial cutoff was 6 \AA. Different lines correspond to different descriptors: Parrinello-Behler (PS), Angular Fourier Series (AFS),  bispectrum (BS) and power spectrum (PS). The two versions of the bispectrum differ in the handling of the radial degrees of freedom.}
  \label{fgr:reconstruct_0216}
\end{figure}

\begin{figure}[h]
  \centering
  \includegraphics[width=4cm]{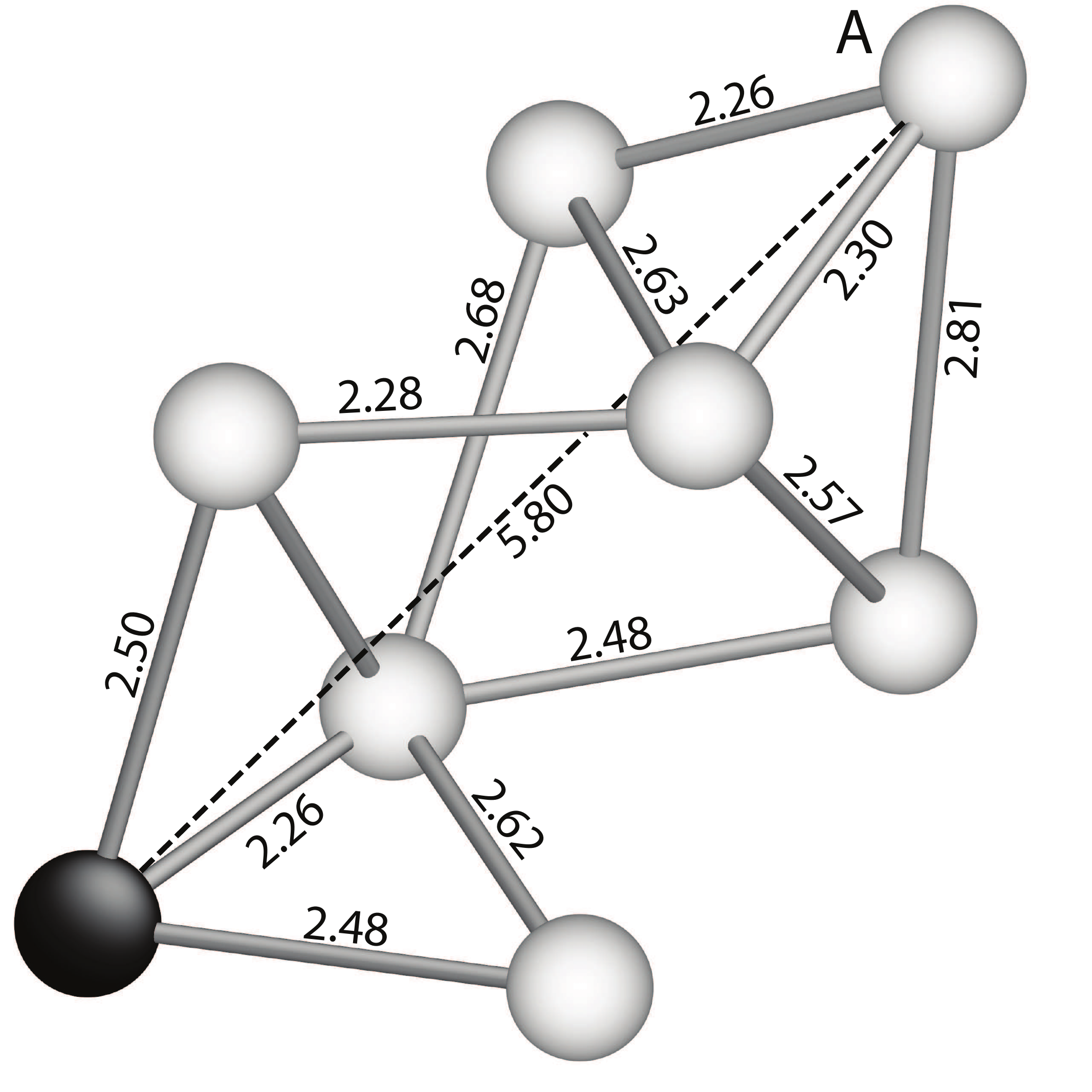}
  \includegraphics[width=4cm]{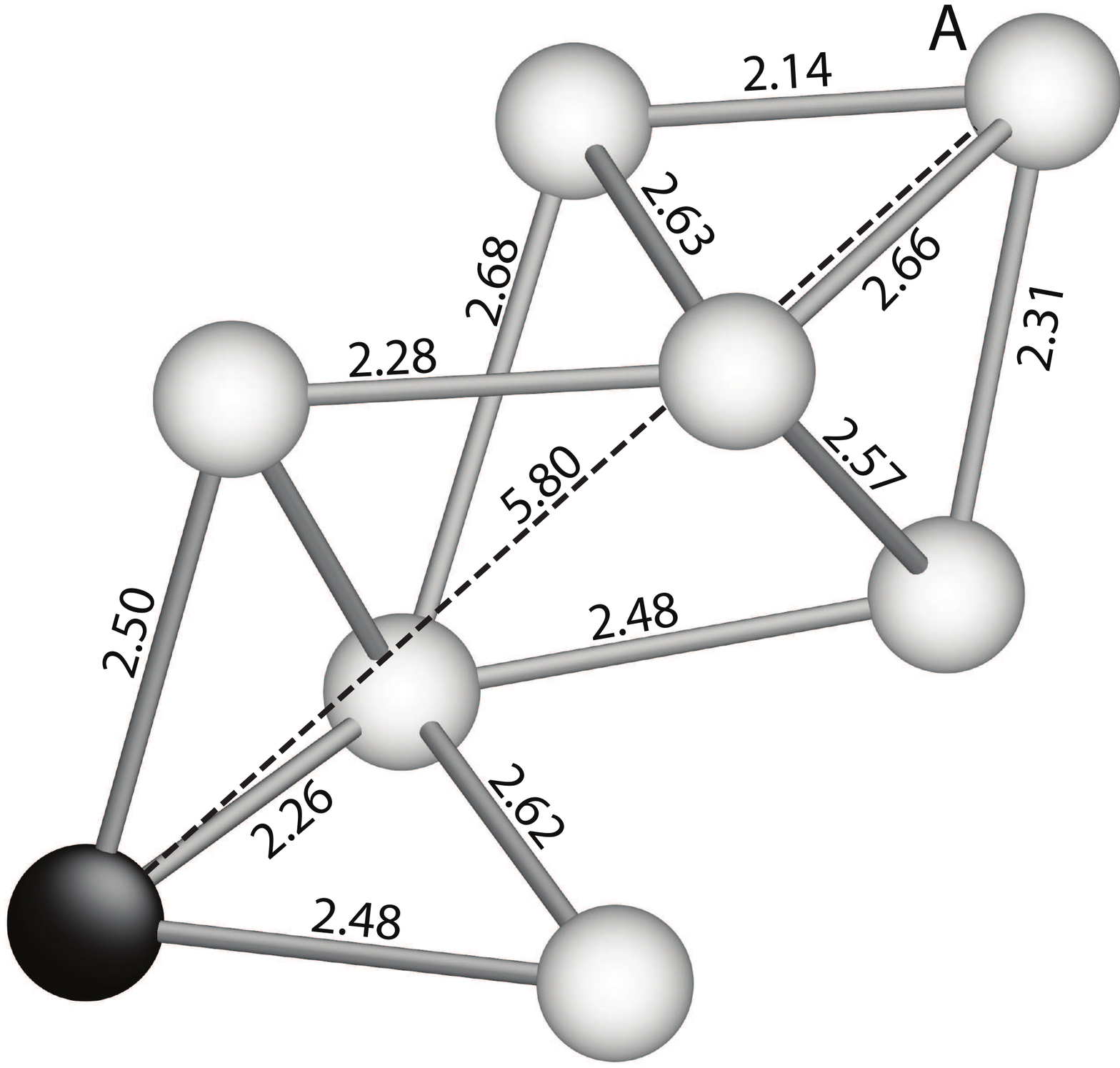}
  \caption{Two $\textrm{Si}_8$ clusters that differ by $d_\textrm{ref} = 0.7 \textrm{\AA}^2$. The reference atom, \ie the centre of the rotation, is coloured black. The only difference between the two clusters is the relative position of the furthest atom, A.}
  \label{fgr:dref_bisp}
\end{figure}

\begin{figure}[h]
  \includegraphics[width=8.6cm]{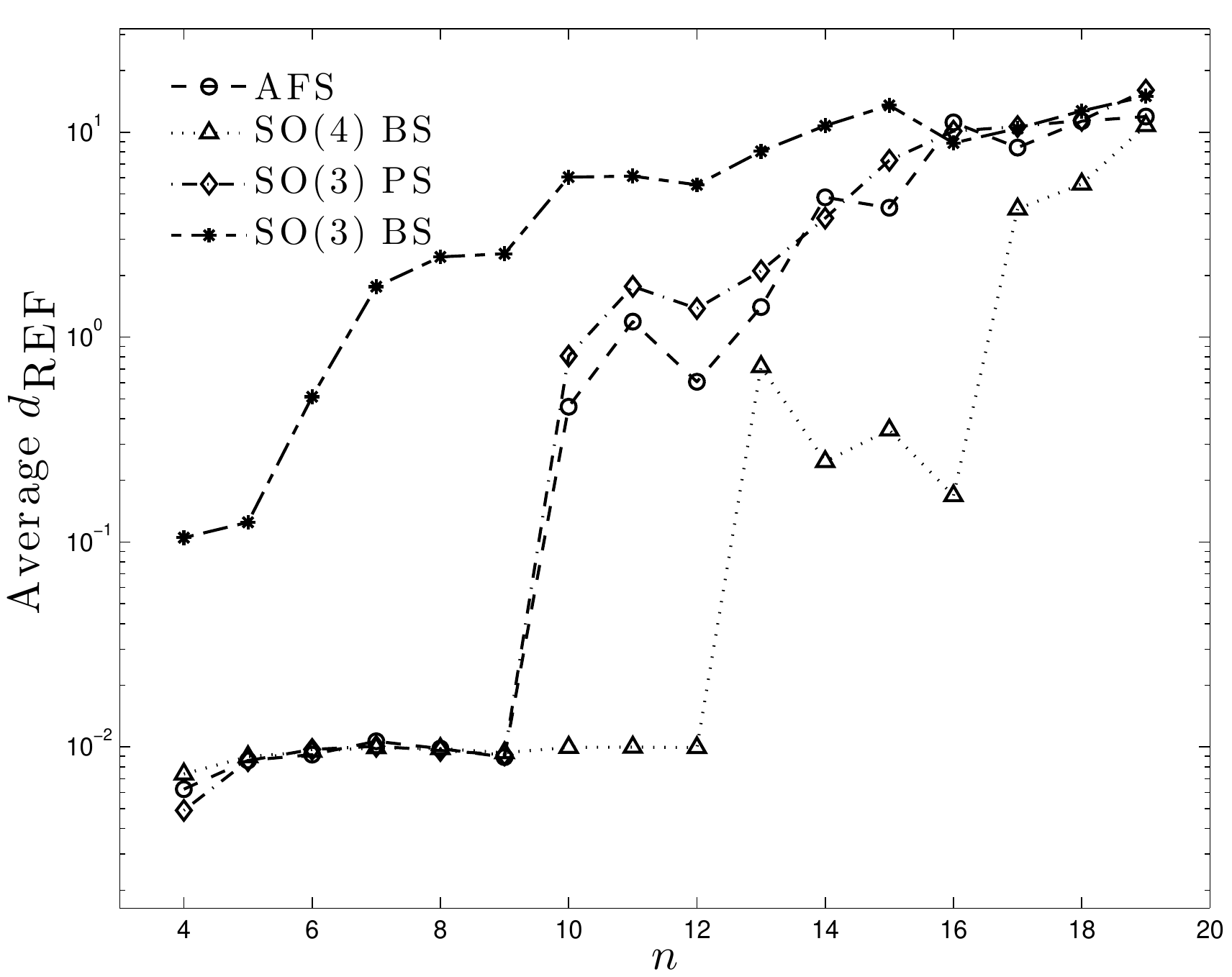}
  \caption{Difference between the target and reconstructed structures after randomisation by $1.6$\AA\ and minimisation, as a function of the number of atoms in the cluster, $n$, averaged over 10 targets for each cluster size.  The cutoff was 9\AA. The line types corresponding to the different descriptors are the same as in Figure~\ref{fgr:reconstruct_0216}.}
  \label{fgr:reconstruct_9A}
\end{figure}

In the first set of reconstruction experiments, in order to provide a fair comparison, the truncation of the formally infinite set of descriptors was chosen in such a way that the finite descriptors had roughly equal numbers of components: 51 in total for the SO(3) bispectrum and PB descriptors and 50 for the AFS and SO(3) power spectrum. This corresponds to a truncation of the SO(4) bispectrum with $2 j_\mathrm{max}=5$ (the factor of 2 on account of the half-integer nature of $j$), the SO(3) bispectrum with $l_\mathrm{max}=4$ and $n_\mathrm{max}=3$, the PB descriptor with its published parameters\cite{Artrith:2012fw} and the AFS and SO(3) power spectrum using $l_\mathrm{max}=9$ and $n_\mathrm{max}=5$. We note that in case of the PB descriptor the band limit of the angular descriptors (corresponding to our $l_\mathrm{max}$ or  $j_\mathrm{max}$) was $\zeta_\mathrm{max}=16$ and only the values $\zeta=1,2,4,16$ are used.

Figure~\ref{fgr:reconstruct_0216} shows the quality of reconstruction for different cluster sizes, based on the PB, AFS, SO(3) power spectrum, SO(3) bispectrum and SO(4) bispectrum as given by the reference distance $d_\textrm{ref}$ achieved, averaged over 10 reconstruction trials for each cluster size $n$. The general trend is the same for all descriptors: as the number of neighbours increases, the average $d_\textrm{ref}$ increases, and thus the faithfulness of the reconstruction decreases. Noting that the stopping criterion for the reconstruction process was $d_\textrm{ref} < 10^{-2}$\AA$^2$, larger randomisation of the initial atomic coordinates (bottom panel) reveals the poor representation power for all descriptors using this parameter set for $n > 10$, and the neighbour configuration becomes impossible to determine from the descriptor. 

The poor quality of representation is partly attributable to the decrease in sensitivity to the positions of atoms near the cutoff. For example, Figure~\ref{fgr:dref_bisp} shows two Si$_8$ clusters for which none of the descriptors lead to perfect reconstructions (resulting in the observed peak on Figure~\ref{fgr:reconstruct_0216}). The atom marked A in the figure is within the 6\AA\ cutoff, but close to it. In order to separate out this effect, we repeated the reconstruction experiments with a radial cutoff of 9\AA\ (omitting the PB descriptor now since there is no published parameter set for this cutoff). The results are shown in Figure~\ref{fgr:reconstruct_9A} for the larger initial randomisation. The peak near $n=8$ is now absent, and the transition from faithful reconstruction (for $n \leq 9$ for the SO(3) power spectrum and AFS, and for $n \leq 12$ for the SO(4) bispectrum) to failure for larger $n$ is much clearer.

\begin{figure}[h]
  \includegraphics[width=8.6cm]{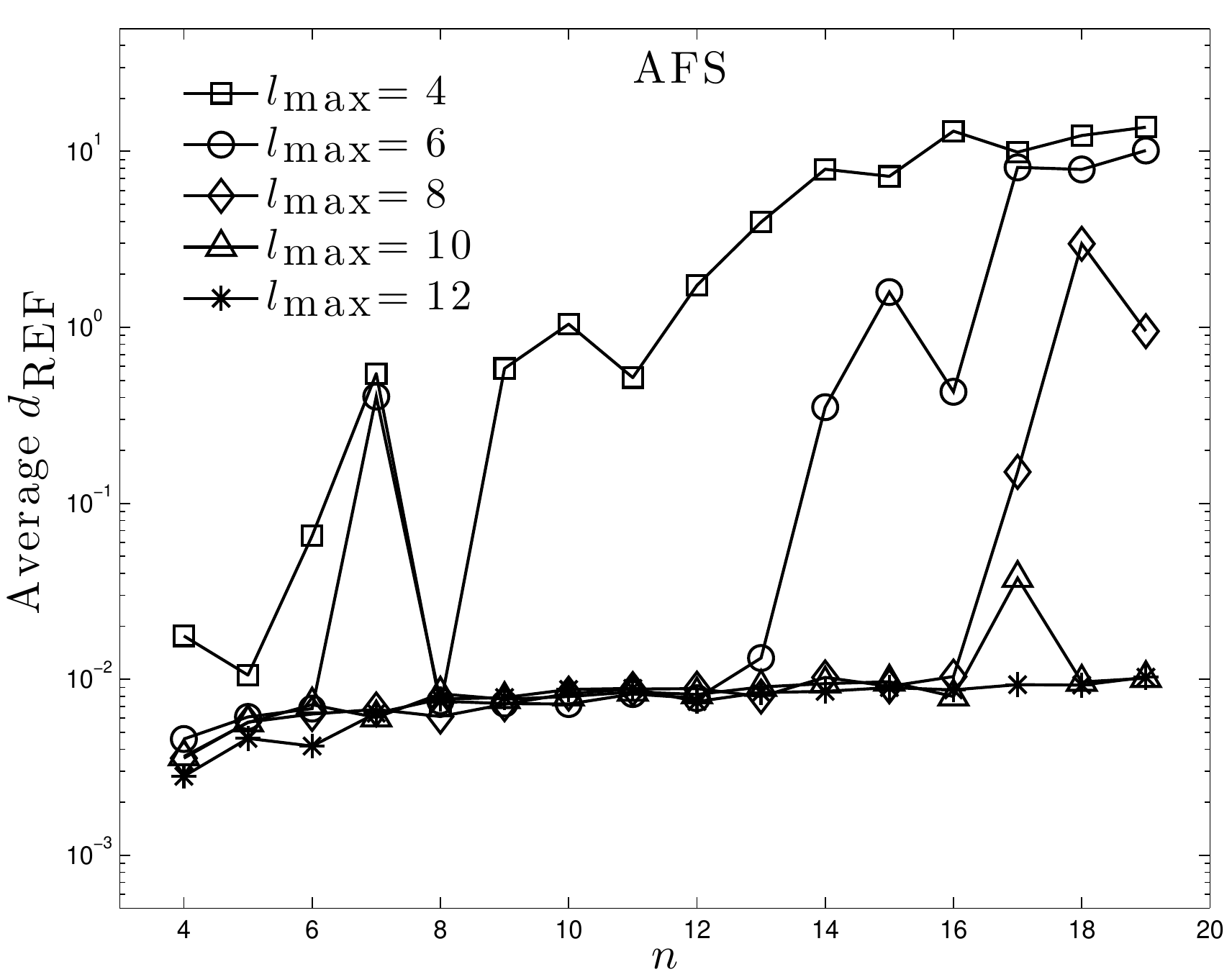}
   \includegraphics[width=8.6cm]{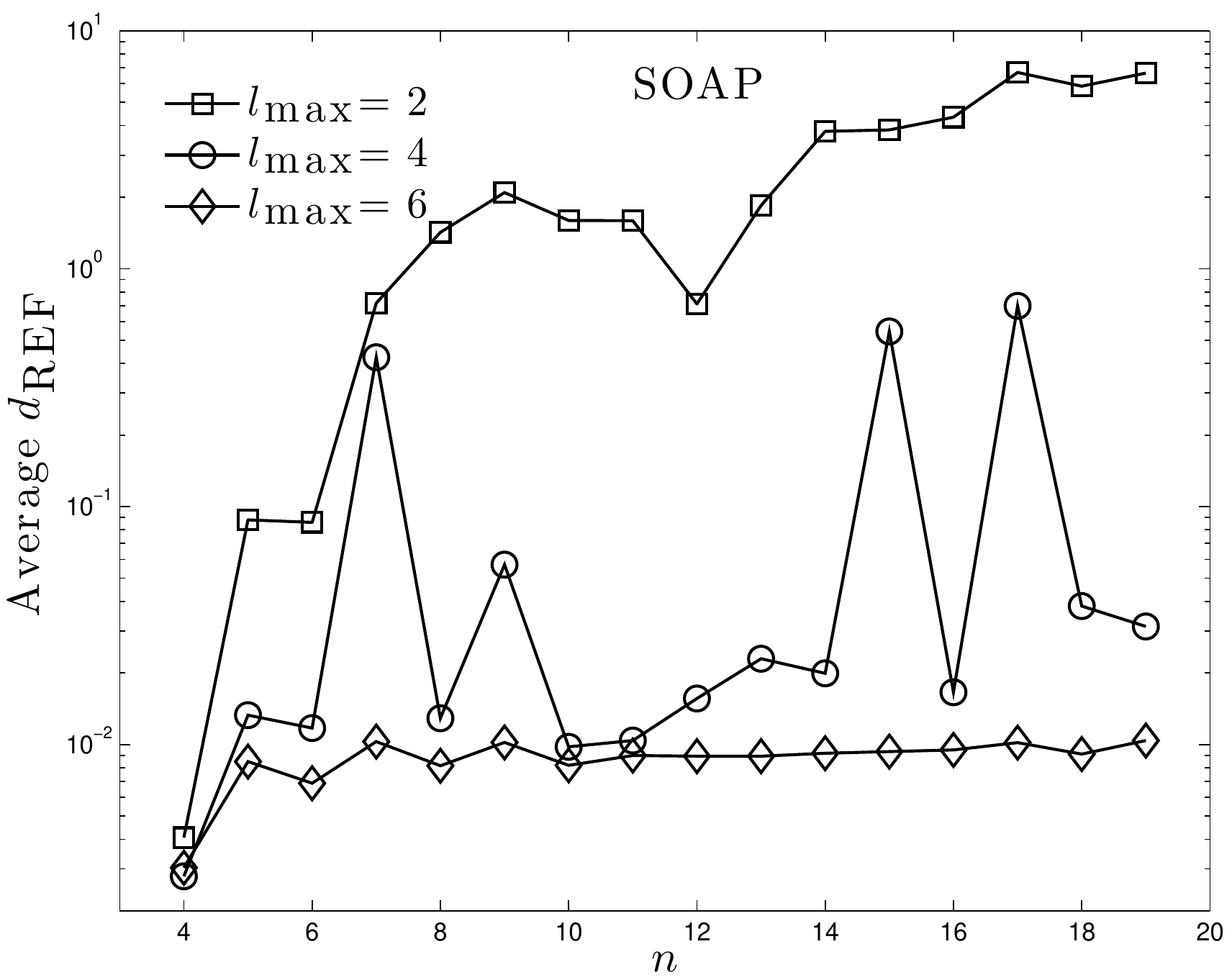}
  \caption{Difference between target and reconstructed structures after randomisation by 1.6\AA\  and minimisation as a function of the number of atoms in the cluster, $n$, using the AFS descriptor (top) and SOAP (bottom) with a radial cutoff of 9\AA, averaged over 10 targets for each cluster size. For the case of AFS, the different curves correspond to different numbers of components of the descriptor, achieved by varying the truncation of the angular expansion, while in the case of SOAP, we varied the truncation of the expansion of the atomic neighbourhood density, which corresponds to varying the accuracy of the evaluation of the similarity measure.}
  \label{fgr:afs_conv}
\end{figure}

Since all the descriptors are likely to be over-complete when the infinite series of the basis set expansion is not truncated,  the reconstruction quality is expected to increase with increasing descriptor length. To verify this, Figure~\ref{fgr:afs_conv} shows the reconstruction quality of the AFS descriptor for varying  truncations of the angular part of the basis set. The representation becomes monotonically better for higher angular resolutions. However, this comes at the price of introducing ever more highly oscillating basis functions, which might be less and less suitable for fitting generally smooth potential energy surfaces.

Figure \ref{fgr:afs_conv} also shows the achieved reference values when using the SOAP similarity measure. In this case, rather than minimising the difference between descriptors, we optimised the candidate structure until its normalised similarity to the target as given by equation \refpar{eq:similarity_kernel} was as close to unity as possible. In contrast to the other descriptors, SOAP with the modest band limit of $l_\mathrm{max}=6$ performs perfectly for all structures, without showing any degradation for larger numbers of neighbours.  

\begin{figure}[htb]
  \includegraphics[width=8.6cm]{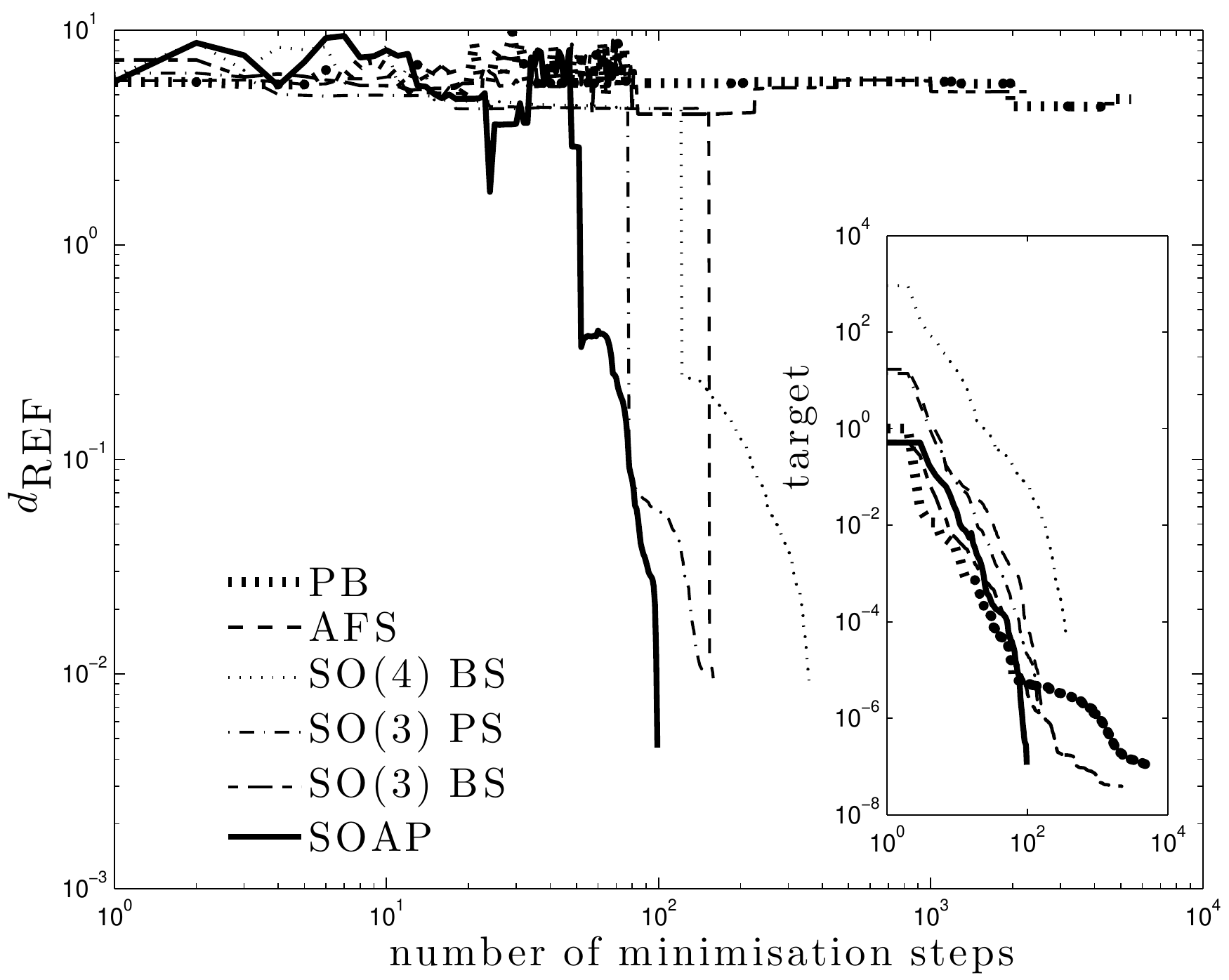}
  \caption{Convergence of reference distance measure during a typical reconstruction procedure. The inset shows the value of the minimisation target approaching zero, \ie descriptor equivalence, for all of the descriptors.}
  \label{fgr:convergence}
\end{figure}

To verify that the above results are not affected by artefacts of the minimisation procedure, \eg getting stuck, Figure~\ref{fgr:convergence} shows the convergence of the reference measure $d_\textrm{ref}$ during a minimisation as well as the corresponding convergence of the target function (the difference between the target and candidate descriptors). There was no difficulty in converging the target function to zero (the global minimum) for any of the complete (or over-complete) descriptors or SOAP, while the reference similarity converged to a non-zero value for incomplete descriptors. 

\subsection{Gaussian Approximation Potentials}
Our main motivation for assessing different approaches to representing atomic neighbour environments is to determine their efficacy for generating interatomic potentials. Therefore, as a final test, we fitted a series of interatomic potentials for $\textrm{Si}_{3-19}$, based on different descriptors, using our Gaussian Approximation Potential (GAP) framework\cite{Bartok:2010fj}. The training and the testing configurations were obtained from tight-binding\cite{Porezag:1995tk} molecular dynamics trajectories run at the temperatures between 500~K and 2000~K. We used four sets of cluster configurations, containing 2000, 4000, 6000 and 8000 atomic environments for the training, corresponding to a total of 180, 360, 540 and 720 unique cluster configurations, respectively. The test set contained 12000 atomic environments, independent from those used in the fitting procedure. 

\begin{figure}[h]
  \includegraphics[width=8.6cm]{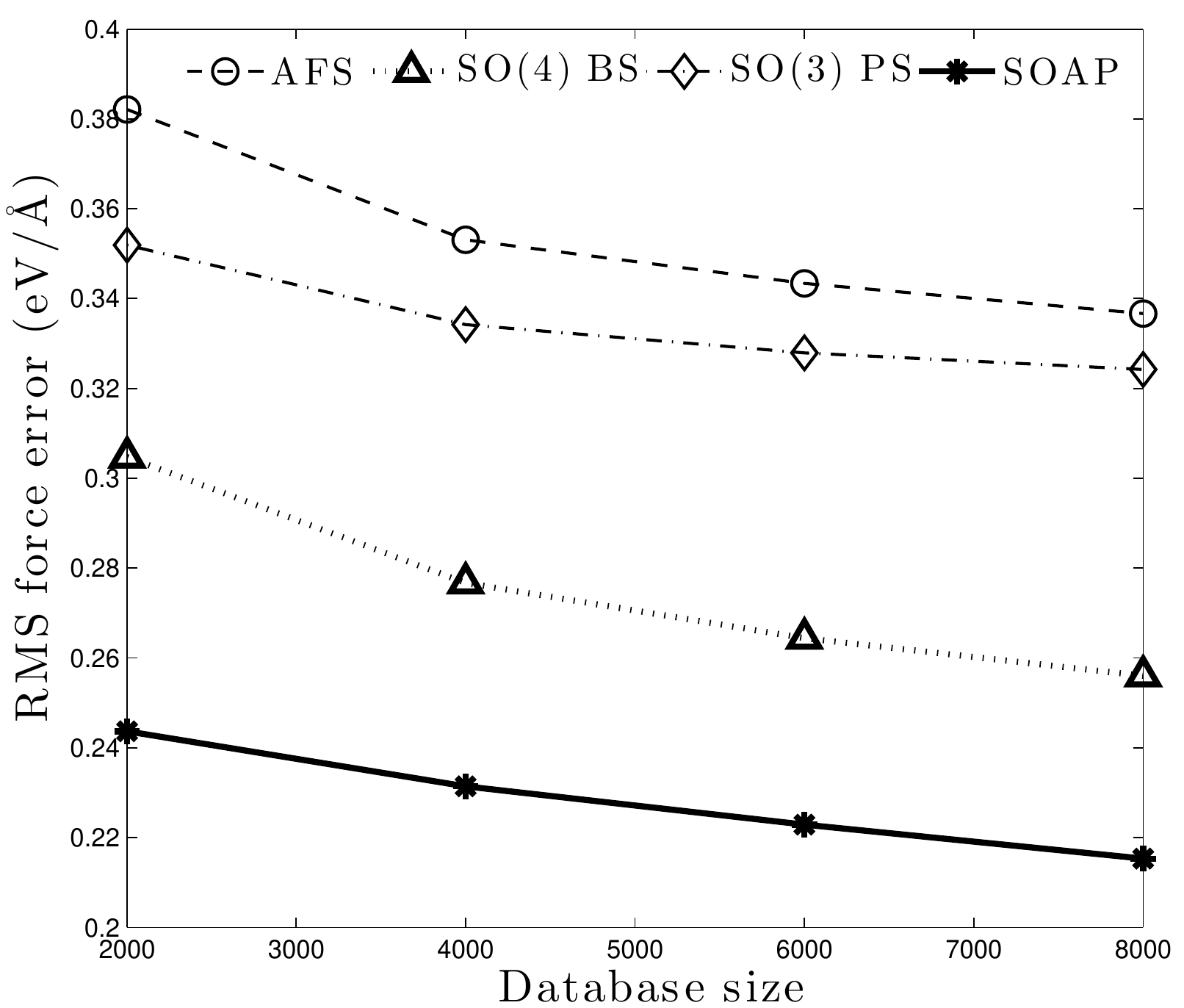}
  \caption{Quality of GAP potentials constructed using different descriptors, as a function of the size of the database used for the fit, with all configurations drawn from Si$_n$ clusters with $3\leq n\leq19$. The angular band limit was $l_\mathrm{max} =12$ in all cases (equivalent to $2 j_\mathrm{max}=12$ for the SO(4) bispectrum).}
  \label{fgr:teach_fcorr}
\end{figure}

We tested AFS, the SO(3) power spectrum and the SO(4) bispectrum using the
squared exponential covariance kernel~\refpar{eq:sekernel} as well as SOAP for
potential fitting. The accuracy of the resulting potential energy surfaces is
shown in Table~\ref{tbl:GAP} as a function of the angular band limit, and in
Figure~\ref{fgr:teach_fcorr} as a function of the database size. Both
demonstrate that SOAP outperforms the other descriptors. As can be expected from the reconstruction tests (\cf Figure~\ref{fgr:afs_conv}) the fit gets better with all descriptors when a larger angular resolution is used, with the error not yet saturated for $l_\mathrm{max}=12$.
Similarly, increasing the database size makes the fit more accurate, and one can expect improvement if even more than 8000 atomic environments are used (this corresponds to, on average, just 40 configurations for each cluster size).

\begin{figure}[h]
  \includegraphics[width=8.6cm]{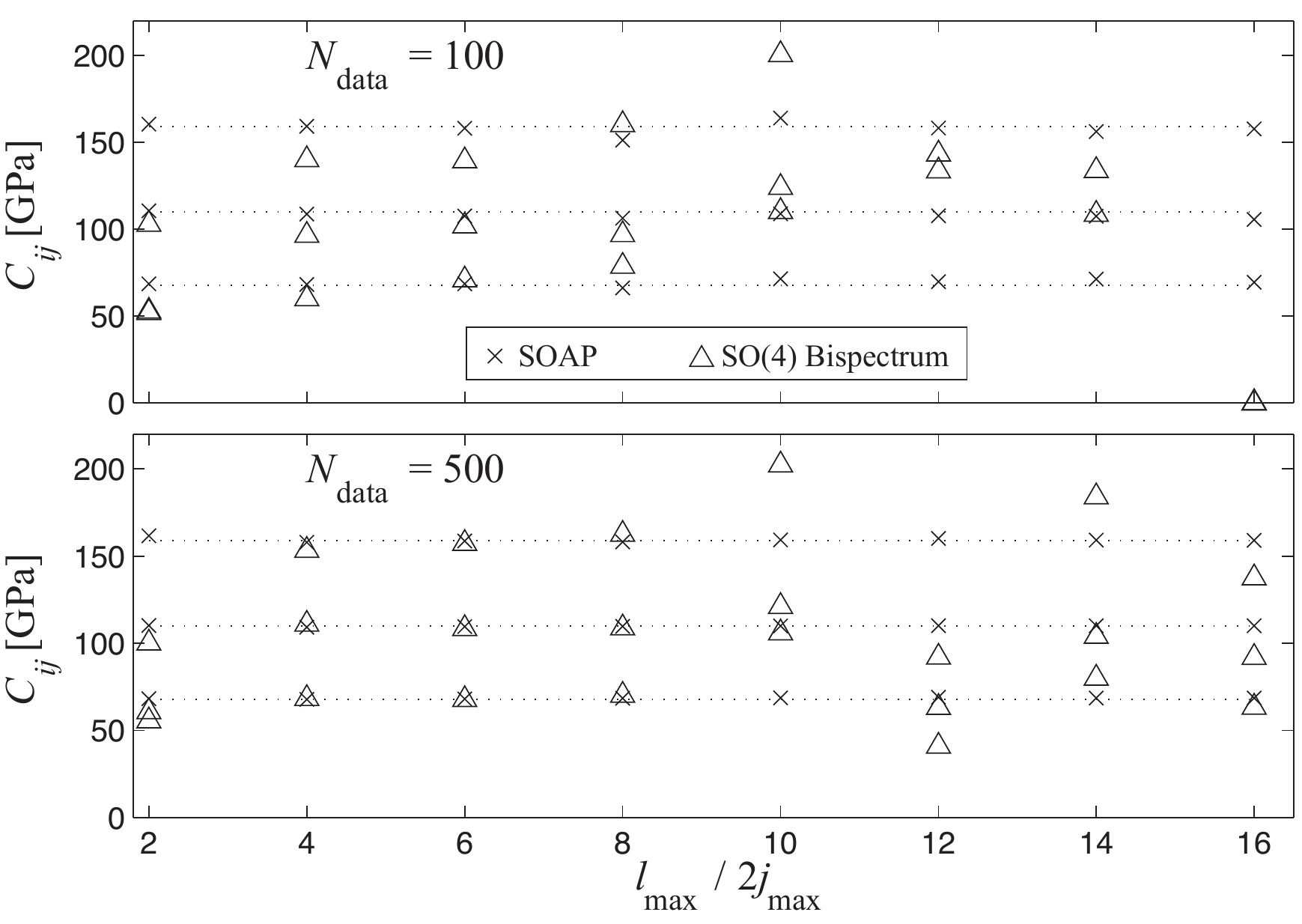}
  \caption{Elastic constants $C_{11}, C_{12}$ and $C^0_{44}$ of GAP models for
bulk silicon using the SO(4) bispectrum and SOAP, as a function of the angular
band limit $l_\mathrm{max}$ (for SOAP) or $j_\mathrm{max}$ (for the bispectrum).
The top and bottom panels correspond to a database size of 100 and 500 configurations. The dashed line indicates exact values of the tight-binding model which was used to generate the database.}
  \label{fgr:cij}
\end{figure}

Perhaps contrary to initial expectations, making all the descriptors more faithful by using a larger angular band limit is not necessarily beneficial. Descriptor components corresponding to high angular momentum channels involve angular basis functions that are highly oscillatory, and can thus degrade the fitted potential energy surface. To demonstrate this, we constructed a GAP model for bulk silicon using a database of configurations with randomly displaced atoms in randomly distorted unit cells, containing two atoms.
Figure~\ref{fgr:cij} shows the elastic constants of the GAP fits as a function of the angular band limit for SOAP and the SO(4) bispectrum descriptor (which performed the best in our reconstruction and cluster PES tests compared to the other descriptors). In case of the bispectrum, the elastic constants of the model improve up to $2 j_\mathrm{max}=8$, but then deteriorate dramatically, irrespective of the  database size. SOAP does not show this behaviour, and leads to reasonable elastic constants using the smaller database, and is already well converged for $l_\mathrm{max}=6$ using the larger database. Given the view of SOAP as an SO(3) power spectrum, it appears that the benefit comes from the combination of building the atomic neighbourhood density from smooth Gaussians and using the dot product kernel --- both directly linked to the construction of SOAP as a smooth similarity measure --- in contrast to using Dirac-delta functions for the density and a squared exponential kernel for the other descriptors.  

%

\begin{table}[h]
\small
\caption{Quality of the GAP potential energy surface using different descriptors and angular band limits, as measured by the RMS energy and force errors. The fitting database contained 8000 atomic neighbourhoods in Si$_n$ clusters with $3\leq n\leq19$. The units of the RMS errors of the energy and force are meV/atom and eV/\AA, respectively.}
\label{tbl:GAP}
\begin{ruledtabular}
  \begin{tabular}{lccc}
 
&Angular band limit & RMS(e) & RMS(f) \\
 \colrule
 & $l_\textrm{max}$ \\
 & 6  & 50.0 & 0.37 \\
 AFS& 8  & 47.0 & 0.35 \\
 $n_\textrm{max}=6$& 10 & 45.7 & 0.34 \\
 & 12 & 44.7 & 0.34 \\
 \colrule
 &
$2 j_\textrm{max}$ \\
&  6  & 27.6 & 0.28 \\
SO(4) BS& 8  & 22.8 & 0.26 \\
$r_0 = \tfrac{4}{3}\:r_\textrm{cut}$& 10 & 20.2 & 0.25 \\
& 12 & 19.2 & 0.26 \\
\colrule
&
$l_\textrm{max}$ \\
 & 6  & 41.5 & 0.36 \\
SO(3) PS& 8  & 37.1 & 0.34 \\
$n_\textrm{max} = 6$& 10 & 35.7 & 0.33 \\
& 12 & 35.0 & 0.32 \\
 \colrule
  &
$l_\textrm{max}$ \\
&  2 & 21.4 & 0.23 \\
SOAP& 4 & 17.6 & 0.21 \\
$\alpha = 2$, $\zeta=4$& 6 & 17.0 & 0.21 \\
& 8 & 15.3 & 0.22 \\
  \end{tabular}
\end{ruledtabular}
\end{table}

%


\section{Conclusion}
In this paper we discussed a number of approaches to representing atomic neighbour environments within a finite cutoff such that the representation is a continuous and differentiable function of the atomic positions and is invariant to global rotations, reflections, and permutations of atoms. We showed that the Steinhardt bond-order parameters  are equivalent to certain elements of the SO(3) angular power spectrum and bispectrum.
To incorporate radial information, and therefore provide a full description of the atomic neighbour environment, we reviewed the construction of the SO(4) power spectrum and bispectrum as an alternative to introducing explicit radial basis functions.  We also demonstrated that all these constructs, as well as the descriptors suggested by Parrinello and Behler, use very similar terms and form part of a general family that is based on the bond angles. In practice, when the expansion is truncated, the faithfulness of the descriptors decreases as the number of neighbours increases, leading to a tuneable trade-off between the size of the descriptor and its faithfulness in terms of its ability to represent the atomic environment uniquely up to symmetries. With typically used parameters, however, the faithfulness of the descriptors are quite different, and all descriptors fail for Si clusters with more than 13 atoms. In order to improve on this, we therefore introduced a similarity measure between atomic neighbour environments called Smooth Overlap of Atomic Positions, which does not suffer from these difficulties and demonstrates excellent faithfulness for any number of neighbours. We also tested the performance of the descriptors for fitting models of small silicon clusters and bulk silicon crystal and found that SOAP leads to a more accurate and much more robust potential energy surface. 

\begin{acknowledgments}
APB gratefully acknowledges funding from Magdalene College, Cambridge. Figures~\ref{fgr:dref} and \ref{fgr:dref_bisp} were generated using AtomEye\cite{Li:2003wy}. This work was partly supported by the European Framework Programme (FP7/2007-2013) under grant agreement no. 229205. 
\end{acknowledgments}


\bibliography{representation_paper}

\end{document}